\documentclass[
superscriptaddress,
reprint,
prb,
floatfix,
]{revtex4-1}

\usepackage[utf8]{inputenc}

\usepackage{graphicx}
\usepackage{dcolumn}
\usepackage{bm}
\usepackage{color}
\usepackage{multirow}
\usepackage{array}
\usepackage{tabularx}
\newcolumntype{K}[1]{>{\centering\arraybackslash}p{#1}}
\usepackage{xspace}  
\usepackage{amsmath}
\usepackage{xcolor}  

\graphicspath{{./}}

\begin{document}

\newcommand{\ecuteps}{\epsilon_{\scriptsize\textrm{cut}}}
\newcommand{\bgw}{\textsc{BGW}\xspace}
\newcommand{\abinit}{\textsc{ABI}\xspace}
\newcommand{\yambo}{\textsc{YMB}\xspace}
\newcommand{\fullbgw}{\textsc{BerkeleyGW}\xspace}
\newcommand{\fullabinit}{\textsc{Abinit}\xspace}
\newcommand{\fullyambo}{\textsc{Yambo}\xspace}
\newcommand{\exx}{E_{\scriptsize\textsc{EXX}}}
\newcommand{\qp}{\textrm{\scriptsize{QP}}}
\newcommand{\ks}{\textrm{\scriptsize{KS}}}
\newcommand{\unfmode}{\omega^{\scriptsize\textrm{unf.}}}
\newcommand{\Neps}{N_{\scriptsize\textrm{eps.}}}
\newcommand{\Nsig}{N_{\scriptsize\textrm{sig.}}}
\newcommand{\Npw}{N_{\scriptsize\textrm{PW}}}

\newcommand{\FHJ}[1]{\textcolor{blue}{[FHJ: #1]}}
\newcommand{\TR}[1]{\textcolor{red}{#1}}
\newcommand{\GKA}[1]{\textcolor{purple}{#1}}

\title{Reproducibility in $G_0W_0$ Calculations for Solids}

\author{Tonatiuh Rangel}
\email{trangel@lbl.gov}
\affiliation{Molecular Foundry, Lawrence Berkeley National Laboratory, Berkeley, California 94720, United States}
\affiliation{Department of Physics, University of California at Berkeley, California 94720, United States}
\author{Mauro Del Ben}
\affiliation{Computational Research Division, Lawrence Berkeley National Laboratory, Berkeley, California 94720, United States}
\author{Daniele Varsano}
\affiliation{Centro S3, CNR-Istituto Nanoscienze, I-41125 Modena, Italy}
\affiliation{European Theoretical Spectroscopy Facility (ETSF)}
\author{Gabriel Antonius}
\affiliation{Department of Physics, University of California at Berkeley, California 94720, United States}
\affiliation{Materials Sciences Division, Lawrence Berkeley National Laboratory, Berkeley, California 94720, United States}
\affiliation{Département de Chimie, Biochimie et Physique, Institut de recherche sur l’hydrogène, Université du Québec à Trois-Rivières, Qc, Canada}
\author{Fabien Bruneval}
\affiliation{DEN, Service de Recherches de Métallurgie Physique, Université Paris-Saclay, CEA, F-91191 Gif-sur-Yvette, France}
\affiliation{Molecular Foundry, Lawrence Berkeley National Laboratory, Berkeley, California 94720, United States}
\affiliation{Materials Sciences Division, Lawrence Berkeley National Laboratory, Berkeley, California 94720, United States}
\author{ Felipe H. {da Jornada}}
\affiliation{Department of Physics, University of California at Berkeley, California 94720, United States}
\affiliation{Materials Sciences Division, Lawrence Berkeley National Laboratory, Berkeley, California 94720, United States}
\author{ Michiel J. {van Setten}}
\affiliation{Institute of Condensed Matter and Nanoscience (IMCN), Universit\'e catholique de Louvain, 1348 Louvain-la-Neuve, Belgium}
\affiliation{European Theoretical Spectroscopy Facility (ETSF)}
\affiliation{IMEC, Kapeldreef 75, 3001 Leuven, Belgium}
\author{Okan K. Orhan}
\affiliation{School of Physics, Trinity College Dublin, The University of Dublin, Dublin 2, Ireland}
\author{ David D. {O'Regan}}
\affiliation{School of Physics, Trinity College Dublin, The University of Dublin, Dublin 2, Ireland}

\author{Andrew Canning}
\affiliation{Computational Research Division, Lawrence Berkeley National Laboratory, Berkeley, California 94720, United States}
\author{Andrea Ferretti}
\affiliation{Centro S3, CNR-Istituto Nanoscienze, I-41125 Modena, Italy}
\affiliation{European Theoretical Spectroscopy Facility (ETSF)}
\author{Andrea Marini}
\affiliation{Istituto di Struttura della Materia of the National Research Council, Via Salaria Km 29.3, I-00016 Montelibretti, Italy}
\affiliation{European Theoretical Spectroscopy Facility (ETSF)}
\author{Gian-Marco Rignanese}
\affiliation{Institute of Condensed Matter and Nanoscience (IMCN), Universit\'e catholique de Louvain, 1348 Louvain-la-Neuve, Belgium}
\affiliation{European Theoretical Spectroscopy Facility (ETSF)}
\author{Jack Deslippe}
\affiliation{NERSC, Lawrence Berkeley National Laboratory, Berkeley, California 94720, United States}
\author{Steven G. Louie}
\affiliation{Department of Physics, University of California at Berkeley, California 94720, United States}
\affiliation{Materials Sciences Division, Lawrence Berkeley National Laboratory, Berkeley, California 94720, United States}
\author{Jeffrey B. Neaton}
\email{jbneaton@lbl.gov}
\affiliation{Molecular Foundry, Lawrence Berkeley National Laboratory, Berkeley, California 94720, United States}
\affiliation{Department of Physics, University of California at Berkeley, California 94720, United States}
\affiliation{Kavli Energy Nanosciences Institute at Berkeley, Berkeley, California 94720, United States}


\begin{abstract}
\textit{Ab initio} many-body perturbation theory within the $GW$ approximation is a Green's function formalism widely used in the calculation of quasiparticle excitation energies of solids.
In what has become an increasingly standard approach, Kohn-Sham eigenenergies, generated from a DFT calculation with a strategically-chosen exchange correlation functional ``starting point'', are used to construct $G$ and $W$, and then perturbatively corrected by the resultant $GW$ self-energy.
In practice, there are several ways to construct the $GW$ self-energy, and these can lead to variations in predicted quasiparticle energies. 
For example, for ZnO and TiO$_2$, reported $GW$ fundamental gaps can vary by more than 1~eV. 
In this work, we address the convergence and key approximations in contemporary $G_0W_0$ calculations, including frequency-integration schemes and  the treatment of the Coulomb divergence in the exact-exchange term.
We study several systems,
	and compare three different $GW$ codes:
	\fullbgw, \fullabinit and \fullyambo.
We demonstrate, for the first time, that the same quasiparticle energies for systems in the condensed phase can be obtained with different codes, and we provide a comprehensive assessment of implementations of the $GW$ approximation.
\end{abstract}

\pacs{Valid PACS appear here}

\maketitle

\section{Introduction}
Quantitative prediction of charged single-particle excitations in otherwise interacting many-particle systems such as solids is a key component of the design and discovery of materials and the fundamental understanding of matter at the atomistic level.
A rigorous formalism for computing such particle-like excitations is many-body perturbation theory, in which electron addition/removal energies are solutions to an effective non-Hermitian single-particle Hamiltonian with a non-local energy-dependent potential, or self-energy operator $\Sigma$.
In the so-called $GW$ method,~\cite{PhysRev.139.A796} the self-energy $\Sigma$ is approximated, to lowest order in the screened Coulomb interaction $W$, as $iGW$, where $G$ is the one-electron Green's function.
In a standard approach, $G$ and $W$ are constructed from a (either regular or generalized~\cite{seidl1996}) Kohn-Sham (KS) eigensystem, computed via density functional theory (DFT), and the KS eigenvalues are corrected perturbatively with a one-shot $G_0 W_0$ self-energy, where the subscript indicates that $G$ and $W$ are not updated self-consistently.
By accounting for the screening of the crystal environment, $GW$ is naturally applicable to solids and has proven quite effective in predicting quasiparticle energies of a wide range of crystals.\cite{hybertsen_first-principles_1985,hybertsen_electron_1986,aryasetiawan_gw_1998,aulbur_quasiparticle_2000,onida_electronic_2002,louie_chapter_2006}
However, because of the complexity, computational cost, and the number of convergence parameters involved, numerical approximations are required in $GW$ calculations, and varying algorithms in different codes can sometimes yield distinct results.

Crystalline silicon is probably the most-studied test-bed solid for $GW$.
Having high crystal symmetry and containing only \emph{sp}-bonded  orbitals, silicon is a relatively-simple system, for which $GW$ within standard approximations yields accurate quasiparticle energies and sizable self-energy corrections.\cite{hybertsen_first-principles_1985,hybertsen_electron_1986}
Transition metals~(TMs) and transition metal oxides~(TMOs),  with localized $d$ or $f$ electrons, present a bigger numerical challenge for $GW$.
When dealing with TMs, care should be taken in the technical details and approximations used within $GW$.
For instance, the convergence criteria,\cite{shih_quasiparticle_2010} and the choice of frequency-integration scheme~\cite{shaltaf_band_2008,kang_quasiparticle_2010,stankovski_$g^0w^0$_2011,miglio_effects_2012,larson_role_2013} and  pseudopotentials~\cite{klimes_predictive_2014} can yield substantially different results.
Several $GW$ works for rutile TiO$_2$ have reported gaps ranging from 3.1 to 4.8 eV,\cite{oshikiri_electronic_2003,kang_quasiparticle_2010,chiodo_self-energy_2010,patrick_gw_2012,malashevich_first-principles_2014,zhang_all-electron_2015} while for ZnO gaps published so far range from 2.6 to 4.5~eV.\cite{shih_quasiparticle_2010,stankovski_$g^0w^0$_2011,friedrich_band_2011,*friedrich_erratum:_2011,berger_efficient_2012,samsonidze_insights_2014,klimes_predictive_2014,zhang_all-electron_2016}
Thanks to advances in computational resources and algorithms, recent works have explored convergence beyond past limits,\cite{shih_quasiparticle_2010,stankovski_$g^0w^0$_2011,friedrich_band_2011,*friedrich_erratum:_2011,berger_efficient_2012,klimes_predictive_2014,zhang_all-electron_2016} and accurate pseudopotentials specific for $GW$ have been proposed,\cite{klimes_predictive_2014,pseudo-dojo} a general agreement on the $GW$ quasiparticle energies with different codes has yet to be perceived as being achieved for the difficult cases. 
The growing popularity of $GW$, the multiple dedicated codes used for $GW$, and the existing challenges and discrepancies encountered when  performing $GW$ on increasingly chemically complex systems, such as TMs and TMOs, make it imperative to have reproducibility of predictions from different $GW$ codes.

In this work, we report the results of a detailed comparison of three different plane-wave-based $GW$ codes, and we find that predictions from these codes can agree very well, under given similarly physically sound approximations. For purposes of assessment, we study the representative solids Si, Au, TiO$_2$, and ZnO with the open-source $GW$ codes \fullabinit (\abinit)~\cite{gonze_cpc2016}, \fullbgw (\bgw),\cite{deslippe_berkeleygw:_2012} and \fullyambo (\yambo).\cite{marini_yambo:_2009} Our benchmark calculations provide a framework for users and developers to document the precision of new applications and methodological improvements, and provides standards for the reproducibility of $GW$ calculations.

\section{The $GW$ method in practice} 
The $GW$ method is an interacting Green's function formalism which accounts for the response of the system to addition or removal of a single electron in an interacting $N$-electrons system, via a non-Hermitian, non-local, and frequency-dependent self-energy operator
\begin{equation}
\Sigma(\mathbf{r}, \mathbf{r'}; \omega)= \frac{i}{2\pi} \int \mathrm{d}\omega' \, 
e^{i\omega' \eta}
G(\mathbf{r},\mathbf{r}';\omega+\omega') W(\mathbf{r}, \mathbf{r'};\omega'),
\label{eq:sigma}
\end{equation}
where $\eta$ is a positive infinitesimal and the bare Coulomb potential $v$ and the inverse of the dielectric matrix $\epsilon^{-1}$ are used to construct the screened Coulomb potential
\begin{equation}
W(\mathbf{r},\mathbf{r}';\omega)
= \int d\mathbf{r}''
\epsilon^{-1}(\mathbf{r},\mathbf{r}'';\omega)
v(\mathbf{r}'',\mathbf{r}').
\label{eq:screening}
\end{equation}

In the so-called one-shot $GW$, also known as $G_0W_0$, the quasiparticle energies $E^\qp$ are solved perturbatively from a mean-field Kohn-Sham~(KS) starting point; that is, $G_0$ and $W_0$ are constructed from the KS mean-field. In this approach, which implicitly assumes the KS wavefunctions $\psi^\ks$ are close to the QP wavefunctions $\psi^\qp$, the QP energy of the $i$th state is given by\cite{hybertsen_first-principles_1985,hybertsen_electron_1986}
\begin{align} 
E^\qp_{i} = E^\ks_{i} + \langle \psi^\ks_{i}  | 
\Sigma(E^\qp_{i}) 
- V_{xc} | \psi^\ks_{i} \rangle
\label{eqn:g0w0}
\end{align}
where $V_{xc}$ is the KS exchange-correlation potential, and
$\Sigma$ is evaluated at the QP energy $E^\qp_i$.
A common approximation is to linearize $\Sigma$ in the QP energy with a first-order Taylor expansion around $E_i^{\ks}$, such that 
\begin{align} 
E^\qp_{i} = E^\ks_{i} + 
 Z_i \, \langle \psi^\ks_{i}  | 
\Sigma(E^\ks_{i}) 
- V_{xc}| \psi^\ks_{i} \rangle,
\label{eqn:g0w0-linearized}
\end{align}
with the renormalization factor
\begin{eqnarray}
Z_i =  \left[ 1 - \langle \psi^{\ks}_i| \left. \frac{\partial \Sigma(\omega)}{\partial \omega} \right|_{\omega=E^{\ks}_i} | \psi^{KS}_i \rangle \right]^{-1} .
\end{eqnarray}
As discussed later, the standard linearlization scheme should be used with care as it can lead to relatively large deviations (up to $0.2$~eV in ZnO) in predicted QP energies.

A source of deviation among $GW$ results with different codes is the numerical integration scheme used to evaluate the {\it frequency depencence} of $\Sigma$ in Eq.~\eqref{eq:sigma}.\cite{miglio_effects_2012,stankovski_$g^0w^0$_2011,kang_quasiparticle_2010,shaltaf_band_2008} 
A common practice to reduce computational cost is to approximate the dielectric function with a single-pole via a generalized plasmon-pole model~(PPM). 
For each set of momentum components ($\mathbf{q},\mathbf{G},\mathbf{G'}$), the inverse dielectric function $\epsilon^{-1}$ in this approximation takes the form
\begin{eqnarray}
\textrm{Im} \ \epsilon^{-1}_{\mathbf{G},\mathbf{G'}}
(\mathbf{q},\omega)
&=& A_{\mathbf{G},\mathbf{G'}}(\mathbf{q}) \times \\
&&\left[ 
\delta\big( \omega-\tilde{\omega}_{\mathbf{G},\mathbf{G}'} (\mathbf{q})\big) 
-
\delta\big( \omega+\tilde{\omega}_{\mathbf{G},\mathbf{G}'} (\mathbf{q})\big) 
\right] \nonumber\\[6pt]
\textrm{Re} \ \epsilon^{-1}_{\mathbf{G},\mathbf{G'}}
(\mathbf{q},\omega)&=&
1  
-\frac{A_{\mathbf{G}\mathbf{G}'}
(\mathbf{q})\,
\tilde{\omega}^2_{\mathbf{G}\mathbf{G}'}(\mathbf{q})
}{\omega^2 - \tilde{\omega}^2_{\mathbf{G},\mathbf{G}'}(\mathbf{q})},
\label{eqn:ppm}
\end{eqnarray}
where 
the matrices $A_{\mathbf{G}\mathbf{G}'}(\mathbf{q})$ and $\tilde{\omega}_{\mathbf{G}\mathbf{G}'}(\mathbf{q})$ are to be determined.~\cite{hybertsen_electron_1986}
In the Hybertsen-Louie~(HL) approach, the PPM parameters are determined from sum rules and by evaluating the dielectric function at $\omega=0$.~\cite{hybertsen_electron_1986}
In the Godby-Needs~(GN) scheme, the parameters are set by calculating $\epsilon^{-1}$ at two frequencies: $\omega=0$ and an imaginary frequency close to the plasma frequency.\cite{godby_metal-insulator_1989}
Both \fullabinit\ and \fullyambo use the PPM-GN scheme as default;  \fullbgw uses a PPM-HL version modified to deal with non-centrosymmetric systems.\cite{zhang_evaluation_1989,deslippe_berkeleygw:_2012}
When calculating $\epsilon(\mathbf{q},\mathbf{q}';\omega=0)$ to find the PPM-HL parameters from Eq.~\eqref{eqn:ppm},  it may happen that the dielectric function cannot be satisfactorily approximated by a single-pole model for certain $(\mathbf{q},\mathbf{G},\mathbf{G}')$ leading to imaginary frequencies $\omega_{\mathbf{G},\mathbf{G}'}(\mathbf{q})$.
Such modes, referred to here as {\it unfulfilled PPM modes} $\unfmode$, are neglected in the original version of the PPM-HL.\cite{hybertsen_electron_1986}
Other treatments of the unfulfilled modes are also possible. 
For example, these frequencies can be given an arbitrary value of $\unfmode=1$~Ha, which was the default behavior in \fullabinit{} and \fullyambo.

Beyond PPMs, it is increasingly standard for $GW$ codes to use {\it full-frequency}~(FF) methods, in which the frequency convolution in Eq.~\eqref{eq:sigma} is evaluated numerically.
A straightforward integration method on the real axis~(FF-RA) is available in codes such as \fullyambo\ and \fullbgw.
However, such an integration of $\Sigma$ in Eq.~\eqref{eq:sigma} presents numerical challenges since $G$ and $W$ possess poles close to the real axis.
To avoid this difficulty, in the full-frequency {\it contour-deformation}~(FF-CD) method, the integration contour in Eq.~\eqref{eq:sigma} is deformed into the complex plane, into a region where the integrand is smooth; the alternative integration path must be supplemented with the residues from the poles of $G$, as explained in detail in Refs.~\onlinecite{lundqvist_single-particle_1968,aryasetiawan_gw_1998,giantomassi_electronic_2011}.
The FF-CD method is available in \fullabinit\ and has been recently implemented into \fullbgw{}.\cite{DelBen2019a,DelBen2019b}
For other FF methods we refer the reader to Refs.~\onlinecite{shishkin_implementation_2006,liu_numerical_2015,zhang_all-electron_2015}.

The self-energy is usually split into a frequency-independent exchange part $\Sigma_x$ and a correlation part $\Sigma_c$, so that $\Sigma(\mathbf{r},\mathbf{r'};\omega)=\Sigma_x(\mathbf{r},\mathbf{r'}) + \Sigma_c(\mathbf{r},\mathbf{r'};\omega)$,\cite{bruneval_exchange_2005} where the matrix element of $\Sigma_x$ between two Bloch states reads:
\begin{equation}
\langle i\mathbf{k}|\Sigma_x | j\mathbf{k} \rangle = -
\sum_{\mathbf{q},\mathbf{G}} v(\mathbf{q}+\mathbf{G}) 
\mathcal{F}_{ij\mathbf{k}}(\mathbf{q+G})
\label{eq:sigx}
\end{equation}
and
\begin{equation}
\mathcal{F}_{ij\mathbf{k}}(\mathbf{q+G})
= \sum_{v\in occ.} 
M_{iv\mathbf{k}}(\mathbf{q}+\mathbf{G}) M^*_{jv\mathbf{k}} (\mathbf{q}+\mathbf{G}).
\end{equation}
Here, $M_{iv\mathbf{k}}=\langle i\mathbf{k}|e^{i(\mathbf{q}+\mathbf{G})\cdot\mathbf{r}} |v\mathbf{k-q}\rangle$ are matrix elements for states $i$ and $v$ at k-point $\mathbf{k}$. The expression for $\Sigma_c$ is given in Ref.~\onlinecite{stankovski_$g^0w^0$_2011}.

The exchange term, also present in the evaluation of Fock exchange for hybrid functionals in DFT, 
features a divergence in the Coulomb potential $v(\mathbf{q}+\mathbf{G})=4\pi e^2/|\mathbf{q}+\mathbf{G}|^2$ as $\mathbf{q} \rightarrow 0$ for $\mathbf{G} = 0$.
Several schemes have been proposed to treat the divergence of the Coulomb term.~\cite{hybertsen_textitab_1987,baroni_textitab_1986,gygi_self-consistent_1986,pick_microscopic_1970,massidda_hartree-fock_1993,carrier_general_2007,spencer_efficient_2008,rozzi2006exact,ismail2006truncation,marini_yambo:_2009,deslippe_berkeleygw:_2012}
For instance, in the {\it spherical-cutoff} technique, the Coulomb interaction is attenuated beyond $R_c$ and $v(0)$ is replaced with $2\pi e^2 R_c^2$, where the sphere of radius $R_c$ has volume equal to that of the unit cell times the number of k-points. \cite{spencer_efficient_2008} 
In \fullabinit by default the Coulomb singularity is approached by an auxiliary-function integration method detailed in Ref.~\onlinecite{carrier_general_2007}.
Other codes avoid the Coulomb singularity by replacing the value of $q \rightarrow 0$ in Eq.~\eqref{eq:sigx} by an integral around $q\simeq 0$.\cite{pulci_ab_1998,marini_yambo:_2009,deslippe_berkeleygw:_2012}
This method is applicable to any $\mathbf{q}$~point in the BZ by assuming,
\begin{eqnarray}
\langle i\mathbf{k}|\Sigma_x | j\mathbf{k} \rangle &=& -
\sum_{\mathbf{q},\mathbf{G}} \int_{R_{\mathbf{q}+\mathbf{G}}}
\frac{d\mathbf{q}'}{\Omega(R_\mathbf{q+G})} \,
v(\mathbf{q}+\mathbf{q}'+\mathbf{G}) 
\nonumber \\[6pt] 
 &\times&
 \mathcal{F}_{ij\mathbf{k}}(\mathbf{q+G}),
\label{eq:sigx2}
\end{eqnarray}
where the integral is performed over the BZ region  $R_\mathbf{q+G}$, which is associated with a volume $\Omega(R_\mathbf{q+G})$, and centered around each $\mathbf{q}+\mathbf{G}$ point. This method gives the effect of a larger sampling of points around $\mathbf{q}$ assuming that $\mathcal{F}(\mathbf{q+G})$ is constant over that region.  

In the ``random integration method'' (RIM) implemented in \fullyambo\cite{marini_yambo:_2009} and ``Monte Carlo averaging'' (MC average) technique used in \fullbgw\cite{deslippe_berkeleygw:_2012} the integral is evaluated using a stochastic scheme. 
In both codes a stochastic scheme is also used to evaluate every term of the form $\int d^n \mathbf{q} f(\mathbf{q}) v(\mathbf {q})$ in $\Sigma_c$, as the scheme can straightforwardly account for integration of arbitrary potentials in regions $R_\mathbf{q+G}$ with arbitrary boundaries.
Moreover, with the MC averaging scheme, the analytical behavior of $W(\mathbf{q}\rightarrow0)$ is also appropriately adjusted depending on whether the system behaves like a metal, semiconductor, or displays a graphene-like linearly vanishing density of states; it is also adjusted based on the dimensionality of the system, as discussed in Ref.~\onlinecite{deslippe_berkeleygw:_2012}.
These stochastic integration methods have shown success in accurately computing the Coulomb singularity and in improving the convergence of $\Sigma$ with respect to k-point sampling.\cite{marini_yambo:_2009,deslippe_berkeleygw:_2012}
To facilitate a complete comparison, we also implemented the MC averaging method into \fullabinit\ for the present work, as will be discussed below.

Aside from the physical model employed for the dielectric matrix and the treatment of the Coulomb divergence, we emphasize that several parameters must be converged in order to achieve meaningful $GW$ results.
Both the calculation of $\epsilon$ and $\Sigma_c$ involve unrestricted sums over bands that are truncated up to $\Neps$ and $\Nsig$, respectively.
Additionally, the codes discussed here use plane-wave basis sets; the number of plane-wave basis functions, $\Npw$, used to evaluate $\epsilon$ and $\Sigma$, is expanded up to an energy-cutoff $\ecuteps$.
These three parameters $\Neps$, $\Nsig$, and $\Npw$ are interdependent, and their convergence needs to be addressed simultaneously.\cite{shih_quasiparticle_2010,stankovski_$g^0w^0$_2011}
Here, we extrapolate the $GW$ QP gaps (energy eigenvalue differences) to the complete basis set~(CBS) limit with a function of the form\cite{DelBen2019b}
\begin{align}
f(\Neps, & \Npw, \Nsig) =  \nonumber \\
 & \left(\frac{a_1}{\Neps} + b_1  \right)
   \left( \frac{a_2 }{\Npw}  + b_2 \right)
   \left(\frac{a_3  }{\Nsig} + b_3 \right),
\label{eqn:extrapolation}
\end{align}
where $a_1$, $a_2$, $a_3$, $b_1$, $b_2$, and $b_3$ are constants to be determined.
Other important convergence parameters and considerations include the k-point sampling of the Brillouin zone, pseudopotential choice, basis used to describe the wavefunctions, and in the case of full-frequency calculations, the frequency sampling on the real and imaginary axis.

\section{Technical details}
\label{sect:tech-dets}
In what follows, we compare $GW$ calculations for several materials using three codes implementing the same approaches.
For all materials considered, we fix the lattice parameters to the experimental values. These are, for Si in the diamond structure, fcc Au, rutile TiO$_2$, and wurtzite ZnO, respectively, 5.43~\AA, 4.08~\AA, ($a = 4.60$, $c = 2.9$)~\AA, and ($a = 3.25$, $c = 5.20$)~\AA.
  We use norm-conserving Fritz-Haber Institute pseudopotentials with 6, 4, 12 and 20 valence electrons for O, Si, Ti and Zn, respectively.
For Au, we use Optimized Norm-Conversving Vanderbilt Pseudoptentials (ONCVP)~\cite{hamann_optimized_2013} with 19 valence electrons. 
We use a Perdew-Burke-Ernzerhof~(PBE)~\cite{perdew_generalized_1996} starting point for $GW$, except for ZnO in which the Local Density Approximation (LDA) is used for the sake of comparison to previous works.
Our DFT calculations use a k-point mesh and a plane-wave energy-cutoff which ensure that the total energies are converged within 50 meV per unit cell.
The k-point mesh is consistent with that for $GW$ calculations, see below;
we use a plane-wave energy cutoff to represent wavefunctions of 40, 88, 300 and 300 Ry for silicon, gold, TiO$_2$ and ZnO, respectively.
The $GW$ parameters are carefully set to converge quasiparticle energies to 0.1~eV;
for silicon, we use a $\Gamma$-centered Monhorst-Pack grid of $12\times12\times12$ k-points, $\ecuteps=20$~Ry and 300 unoccupied states;
for gold, we use a mesh of $16\times16\times16$ k-points, $\ecuteps=32$~Ry, and 400 unoccupied states;
for rutile TiO$_2$, we use a shifted k-grid of $6\times6\times10$ k-points and the number of unoccupied states and $\ecuteps$ value were extrapolated to the CBS, as detailed in the supplemental materials (SI); and for wurtzite ZnO, we use a shifted k-grid of $8\times8\times5$ k-points, and the unoccupied states and $\ecuteps$ are also extrapolated to the CBS.
We summarize in Table I of the Supplementary Information (SI) all convergence parameters used for tables and figures in this manuscript.

\section{Results and discussion}

\subsection{Silicon}
\label{silicon}

\begin{figure}[th]
\includegraphics[scale=0.95]{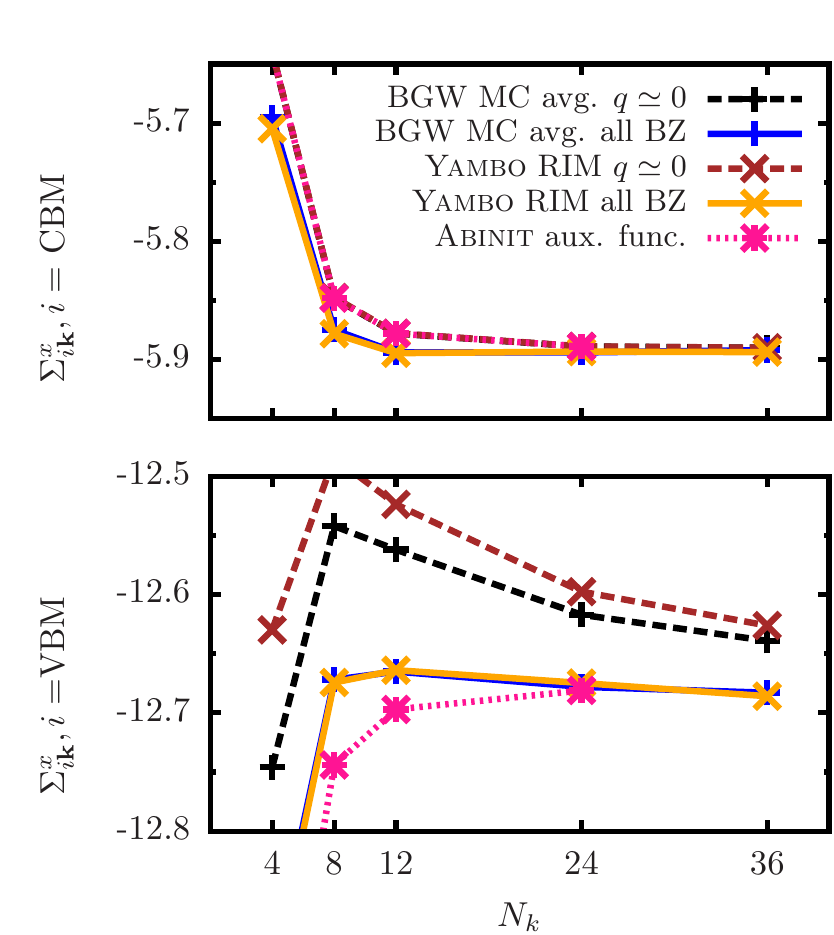}
\caption{
Convergence of the 
matrix elements of $\Sigma_x$ for the VBM and CBM at the $\Gamma$ point for silicon,
with respect to the number of k-points $N_k \times N_k \times N_k$.
In the different codes, several techniques are used to treat the Coulomb singularity (see text). 
}
\label{fig:sigx-silicon}
\end{figure}

\begin{table*}[t]
\begin{tabular}{lcccccccccc}
\hline\hline 
  & \multicolumn{10}{c}{QP energies of silicon (eV)} \\ 
\hline 
\hspace{1.5 cm} & \multicolumn{3}{c}{PPM-GN}
& \hspace{0.2 cm}
& \multicolumn{2}{c}{PPM-HL}  
& \hspace{0.2 cm} 
& \multicolumn{2}{c}{FF-CD} & \hspace{0.1 cm}  FF-RA \hspace{0.1 cm} \\
& \abinit & \bgw & \yambo&& \abinit & \bgw && \abinit & \bgw & \yambo \\
VBM & -0.64 & -0.64 & -0.64 && -0.95 & -0.95 && -0.74 & -0.79 & -0.72\\
CBM & 0.52 & 0.53 & 0.52 && 0.29 & 0.28 && 0.48 & 0.49 & 0.49\\
Gap & 1.16 & 1.17 & 1.16 && 1.24 & 1.24  && 1.22 & 1.28 & 1.21\\[6pt]
\hline\hline
\end{tabular}
\caption{VBM, CBM and fundamental energy-gap of silicon calculated within $GW$ with several codes using different frequency-integration schemes.
Band energies are shown with respect to the DFT VBM.
}
\label{table:silicon-converged}
\end{table*}
We calculate the $GW$ quasiparticle corrections to the bandstructure of bulk silicon, a typical system for $GW$ calculations.
We use a common pseudopotential for all $GW$ calculations, as defined in Section~\ref{sect:tech-dets}.
The effect of the pseudopotential approximation for silicon is discussed in Ref.~\onlinecite{gomez-abal_influence_2008}.

We first study the accuracy of common approximations to treat the Coulomb divergence, which influences the rate of convergence with respect to k-points.
In Fig.~\ref{fig:sigx-silicon}, we show the convergence of the matrix elements of $\Sigma_x$ for the valence band maximum~(VBM) and conduction band minimum~(CBM) at $\Gamma$. 
We consider different techniques to treat the Coulomb singularity, in particular the MC average in \fullbgw for only $\mathbf{q=G}=0$ (black lines, default up to version 1.1 of \fullbgw) and for all $\mathbf{G}$ vectors and q-points in the BZ (blue lines, default starting from version 1.2); the RIM for $q=0$ only (brown lines) and all BZ (orange line) in \fullyambo\; and the auxiliary-function treatment~\cite{carrier_general_2007} in \fullabinit\ (pink lines). 
As expected, both the convergence rate with respect to k-points and the converged number of k-points can differ with the choice of method to treat the Coulomb singularity.
In this case the RIM and MC average approaches converge fastest, with a grid of $8\times8\times8$ k-points being sufficient to converge the $\Sigma_x$ matrix elements for the VBM and CBM within 0.05 eV.

In Table~\ref{table:silicon-converged}, we show converged $G_0W_0$@PBE QP energies for bulk silicon using two different frequency integration schemes and different $GW$ codes.
In fact, we find the same QP energies within 0.05~eV for all codes considered here. 
With respect to the frequency-integration schemes, we find that the PPM in the GN or HL fashions provide a gap for Si within 0.1~eV with respect to the full frequency (FF-CD) reference.
Importantly, for a given frequency-integration scheme, the QP energies obtained with the different codes considered here agree within a tolerance better than 0.05~eV, demonstrating that the same $GW$ corrections can be found with different codes.

We highlight that the VBM, CBM, and gap energies calculated with \fullbgw and \fullabinit\ with FF-CD agree with the energies obtained with \fullyambo\ and FF-RA. This result serves as a numerical verification of the equivalence between the implemented FF-CD and FF-RA integration schemes, which was demonstrated exactly only for the electron gas.~\cite{lundqvist_single-particle_1968}

\subsection{Gold}
\label{sect:gold}
\begin{figure}[h]
\includegraphics[scale=0.95]{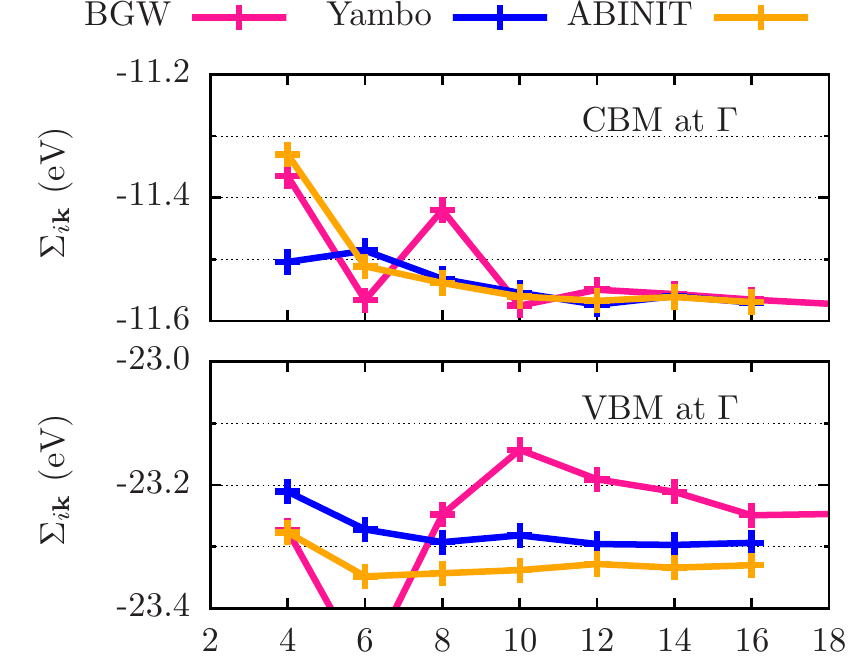}
\caption{Convergence of the $GW$ self-energy of gold.
We show $\Sigma_{i\mathbf{k}}$ matrix elements for $\mathbf{k}=\Gamma$ and $i=$VBM/CBM.
We consider uniform k-point grids of $N_k\times N_k \times N_k$ k points.
The codes used here implement particular sets of approximations to treat metals (see text).
}
\label{fig:gold-k-points}
\end{figure}

We now revisit the $G_0W_0$ corrections to the scalar-relativistic band structure of bulk gold, a relatively difficult case for $GW$ due to convergence issues, the non-negligible influence of semicore orbitals on the band structure, and relativistic effects.~\cite{rangel_band_2012,bernardi_theory_2015}
In what follows, we neglect spin-orbit interactions.
We first converge the number of bands and $\ecuteps$, as detailed in the SI; 
400 unoccupied states and $\ecuteps=32$~Ry ensures a convergence of 0.15~eV in the QP gaps between occupied and unoccupied bands across the Brillouin zone in a relatively large window of energies up to $\sim 15$~eV above the Fermi level. Secondly, we uniformly increase the k-point mesh up to $16 \times 16 \times 16$. We observe differences in $k$-point convergence rate that can be traced to the specific numerical methods used. \fullbgw uses a zero-temperature formalism, and a long wavelength limit of the head ($\mathbf{G=G'}=0$ component) of the inverse dielectric matrix is $\epsilon^{-1}_{00}(q\rightarrow0)\sim q^2$ specific to metals. This in turn modifies the MC averaging scheme, since the head of the screened Coulomb potential $W_{00}(\mathbf{q})$ is now a finite and smooth function for $q\rightarrow0$~\cite{deslippe_berkeleygw:_2012}. 
On the other hand, \fullabinit and \fullyambo use finite-temperature occupation factors, requiring a smearing parameter.
Here we use Gaussian smearing with a broadening of 0.010 Ry.

In Fig.~\ref{fig:gold-k-points}, we show the matrix elements of $\Sigma$ calculated with sets of k-points of increasing size; 
here we set $\ecuteps=32$~Ry and $N=400$.
As mentioned, the rate of convergence depends on the treatment of band occupations and the Coulomb singularity.
While \fullabinit\ and \fullyambo\ use partial occupations consistent with the underlying DFT code, \fullbgw\ uses a zero-temperature scheme where the bands are either fully-occupied or fully-empty.
Moreover, \fullbgw\ uses a particular metal-screening scheme to treat $\epsilon(q\rightarrow0)$ as described in Ref.~\onlinecite{deslippe_berkeleygw:_2012}.
With these different approaches, as expected, the self-energy can converge at different rates with respect to the k-point sampling (see Fig.~\ref{fig:gold-k-points}).
Importantly, when using a relatively-dense mesh of $16\times16\times16$ k-points, the codes considered here agree within 0.1~eV in the predicted self-energy of the VBM/CBM at $\Gamma$, demonstrating that for metals the codes predict the same QP energies when convergence is reached.

In Table~\ref{table:au-gw} we show that the matrix elements of $\Sigma$ for bands around the Fermi level calculated with the different codes.
The scalar-relativistic DFT band structure and the Brillouin zone are shown in the SI.
The $GW$ corrections agree within 0.05~eV, corroborating that at convergence different codes give the same QP energies.

\begin{table}[h]
\begin{tabular}{lccc}
\hline \hline
\hspace{1.5 cm} & \multicolumn{3}{c}{$GW$-PPM self-energy for gold (eV)}\\ 
& \hspace{0.3 cm} \fullabinit \hspace{0.3 cm} & \fullbgw & \hspace{0.3 cm} \fullyambo \hspace{0.3 cm} \\
\hline 
$\Gamma_{12}$ & -23.33 & -23.35 & -23.29\\
$X_5$ & -24.25 & -24.20 & -24.20\\
$X_{4'}$ & -12.98 & -13.08 & -12.97 \\
\hline \hline
\end{tabular}
\caption{Absolute $GW$ self-energy for gold at high-symmetry k points, obtained from a scalar-relativistic PBE DFT calculation. Calculations were performed with three different codes and with the PPM-GN.
}
\label{table:au-gw}
\end{table}

\subsection{Rutile TiO$_2$}

\begin{table*}[th]
\begin{ruledtabular}
\begin{tabular}{lK{1.2cm}K{1.3cm}K{1.3cm}@{\hskip 0.2 cm}K{1.2cm}K{1.3cm}@{\hskip 0.2 cm}K{1.2cm}K{1.3cm}@{\hskip 0.2 cm}K{1.3cm}K{1.3cm}c}
& \multicolumn{10}{c}{Rutile: (unconverged) QP energies obtained with a spherical-cutoff method (eV) }\\
\hline
\hspace{1.2 cm} & \multicolumn{3}{c}{PPM-GN} &
\multicolumn{2}{c}{PPM-HL$^*$} &
\multicolumn{2}{c}{PPM-HL$^\dagger$} &
\multicolumn{2}{c}{FF-CD}& FF-RA \\
& \bgw & \fullabinit & \fullyambo & \bgw & \fullabinit & \bgw & \fullabinit  & \bgw & \fullabinit & \fullyambo\\
VBM & 1.66 & 1.66 & 1.66 & 1.53 & 1.58 & 1.27 & 1.32 & 1.59 & 1.59 & 1.59\\
CBM & 5.47 & 5.47 & 5.47 & 5.62 & 5.58 & 5.98 & 5.94 & 5.45 & 5.45 & 5.43\\
Gap & 3.81 & 3.81 & 3.81 & 4.09 & 4.00 & 4.71 & 4.62 & 3.86 & 3.86 & 3.84\\
\end{tabular}
\end{ruledtabular}
\caption{
QP energies for rutile within a spherical-cutoff technique. 
This comparison is performed with small convergence parameters:
a $6\times6\times10$ k-point grid and $\epsilon_{\text{cut}} = 20$ Ry.
The actual QP energies of rutile are shown in Table.~\ref{table:rutile-gaps}.
We use different codes and frequency-integration schemes (see text)   
For PPM-HL, unfulfilled PPM modes ($\unfmode$) are either $*$ set to $1$~hartree or $\dagger$ neglected. 
} 
\label{table:rutile-cutoff}
\end{table*}

\begin{table}[th]
\begin{ruledtabular}
\begin{tabular}{ccccc}
\multicolumn{5}{c}{Rutile TiO$_2$ QP bandgap (eV)}\\
Code & Potential	& Freq. & $E_g$	&	Ref. \\
\hline
	\fullyambo  & NC-PP & PPM-GN    & 3.2& This work\\
	\fullabinit & NC-PP & PPM-GN    & 3.2& This work\\
           \bgw & NC-PP & PPM-GN    & 3.2 & This work\\
	       \bgw & NC-PP & FF-CD     & 3.3& This work\\
           \bgw & NC-PP	& PPM-HL &3.1	& \onlinecite{malashevich_first-principles_2014} \\
        \textsc{Tombo}	& AE	& PPM-HL& 4.0	& \onlinecite{zhang_all-electron_2015}, \onlinecite{zhang_all-electron_2016}\\
		 \fullyambo	& NC-PP	& PPM-GN &3.6 	&	\onlinecite{chiodo_self-energy_2010} \\
		 \textsc{SaX} 		& NC-PP	& PPM-GN &3.4 	&	\onlinecite{patrick_gw_2012} \\
%
 & AE	&  &	4.8	& \onlinecite{oshikiri_electronic_2003}\\
        \fullyambo	& NC-PP	&  FF-CD	& 3.3	& \onlinecite{kang_quasiparticle_2010}\\
		\textsc{Tombo}	& AE	& FF$^*$& 3.3	& \onlinecite{zhang_all-electron_2015}, \onlinecite{zhang_all-electron_2016}\\
\end{tabular}
\end{ruledtabular}
\caption{We show the fundamental energy-gap of rutile calculated with $G_0W_0$ using different set of approximations within different codes, such as the frequency-integration scheme, basis set and pseudopotentials/all-electron. 
$^*$ FF method in the complex plane\cite{zhang_all-electron_2015}.
}
\label{table:rutile-gaps}
\end{table}

Rutile has been the subject of several $GW$ studies, and the reported $G_0W_0$ gaps range from 3.1 to 4.8~eV~\cite{oshikiri_electronic_2003,chiodo_self-energy_2010,kang_quasiparticle_2010,patrick_gw_2012,malashevich_first-principles_2014,zhang_all-electron_2015,zhang_all-electron_2016}. 
Part of the reported disagreement comes from the treatment of the frequency dependence of $\Sigma$.
As detailed in Ref.~\onlinecite{kang_quasiparticle_2010}, the fundamental gap calculated with certain PPMs can deviate considerably (by up to 1.1~eV) from a full-frequency reference.
The sensitivity of the TiO$_2$ gap to the manner in which the frequency dependence of $\Sigma$ is treated makes rutile an interesting case to investigate the effect and accuracy of PPM and FF methods.
As mentioned previously, we use FHI-type pseudopotentials including semicore states consistently in all calculations performed with different codes. Although the choice of pseudopotentials for $GW$ is not studied in this work, we found that our results for rutile are somewhat modified (by less than 0.1~eV) relative to those obtained with other PPs, such as Gaussian\cite{hartwigsen_relativistic_1998} and pseudo-dojo-v0.2\cite{pseudo-dojo} PPs (see Appendix~\ref{appendix:pps} for more details).

We first examine the $G_0W_0$@PBE QP energies of rutile TiO$_2$ obtained from different codes, frequency-integration schemes, and in the case of PPMs, choices for $\unfmode$, as shown in Table~\ref{table:rutile-cutoff}.
The PPM-GN predicts the VBM, CBM, and gap of rutile within $0.1$~eV of the FF reference.
The accurate performance of the PPM-GN has been observed consistently for other systems, including other transition metal oxides~\cite{shaltaf_band_2008,stankovski_$g^0w^0$_2011,miglio_effects_2012}.

We now examine the PPM-HL and in particular the effect of the different choices for $\unfmode$.
Interestingly, when the terms with unfulfilled PPM modes are set to $1$-Ha, the PPM-HL yields results within $0.1$~eV of the PPM-GN and FF approaches, and when neglecting components with $\unfmode$ the results tend to deviate by up to $0.8$~eV from the FF reference. This clearly indicates that the performance of PPMs for rutile is highly sensitive to the treatment of  unfulfilled PPM modes.
For rutile, $\unfmode$ make up an alarming proportion of the dielectric function ($\sim$54\% of the matrix elements), which suggests the need for a full-frequency treatment of $\epsilon$, in agreement with Ref.~\onlinecite{kang_quasiparticle_2010}.
The fraction of unfulfilled PPM modes is therefore an important indicator of whether a full frequency approach is required.

We now compare the $G_0W_0$ self-energy calculated with different codes in Table~\ref{table:rutile-cutoff}.
When using the PPM-HL, the self-energy can deviate by up to $0.1$~eV for the different codes used here, due to different variants of the PPM-HL being implemented;
while \fullabinit\ implements the original version of PPM-HL in Ref.~\onlinecite{hybertsen_electron_1986}, \fullbgw\ uses a modified version of the PPM to deal with non-centrosymmetric systems as detailed in Ref.~\onlinecite{deslippe_berkeleygw:_2012}.
Assessing these small variations in the PPM is beyond the scope of this work.
When using the PPM-GN or FF methods, the agreement is better than 20 meV, similar to the silicon case.
Importantly, we find that the quasiparticle energies predicted by the different codes agree within $0.1$~eV when using the same treatment of the frequency-dependence.

To converge the $GW$ gap of rutile we extrapolate the interdependent $GW$ parameters ($\ecuteps$, $N_{\text{sig.}}$ and $N_{\text{eps.}}$) to the CBS limit, as described above and in the SI.
The converged bandgap is 3.3~eV for the different codes used here; this result also agrees with previous full-frequency calculations of Refs.~\onlinecite{kang_quasiparticle_2010,zhang_all-electron_2015,zhang_all-electron_2016}, as reported in Table~\ref{table:rutile-gaps}.

\subsection{Wurtzite Zinc Oxide}

\begin{table}[t]
\begin{ruledtabular}
\begin{tabular}{lK{1.1cm}K{1.1cm}K{1.1cm}@{\hskip 0.1 in}K{1.1cm}K{1.1cm}K{1.5cm}}
& \multicolumn{6}{c}{ZnO QP energies (unconverged) (eV) }\\
\hline
& \multicolumn{3}{c}{PPM-GN} &
\multicolumn{2}{c}{FF-CD}& FF-RA \\
& \bgw & \fullabinit & \fullyambo & \bgw & \fullabinit & \fullyambo\\
VBM & 4.26 & 4.29 & 4.26 & 4.27 & 4.26 & 4.26\\
CBM & 8.43 & 8.43 & 8.43 & 8.40 & 8.42 & 8.41\\
Gap & 4.17 & 4.14 & 4.18 & 4.14 & 4.15 & 4.15\\
\end{tabular}
\end{ruledtabular}
\caption{
$GW$ quasiparticle energies of ZnO within a spherical-cutoff technique.
The three $GW$ codes, \fullabinit, \fullyambo\ and \bgw, agree for the calculated QP energies.
This comparison is performed with under-converged parameters:
a $5\times5\times4$ k-point grid and $\epsilon_{\text{cut}}=30$~Ry.}
\label{table:ZnO-spherical-cutoff}
\end{table}

\begin{figure}[h]
\includegraphics{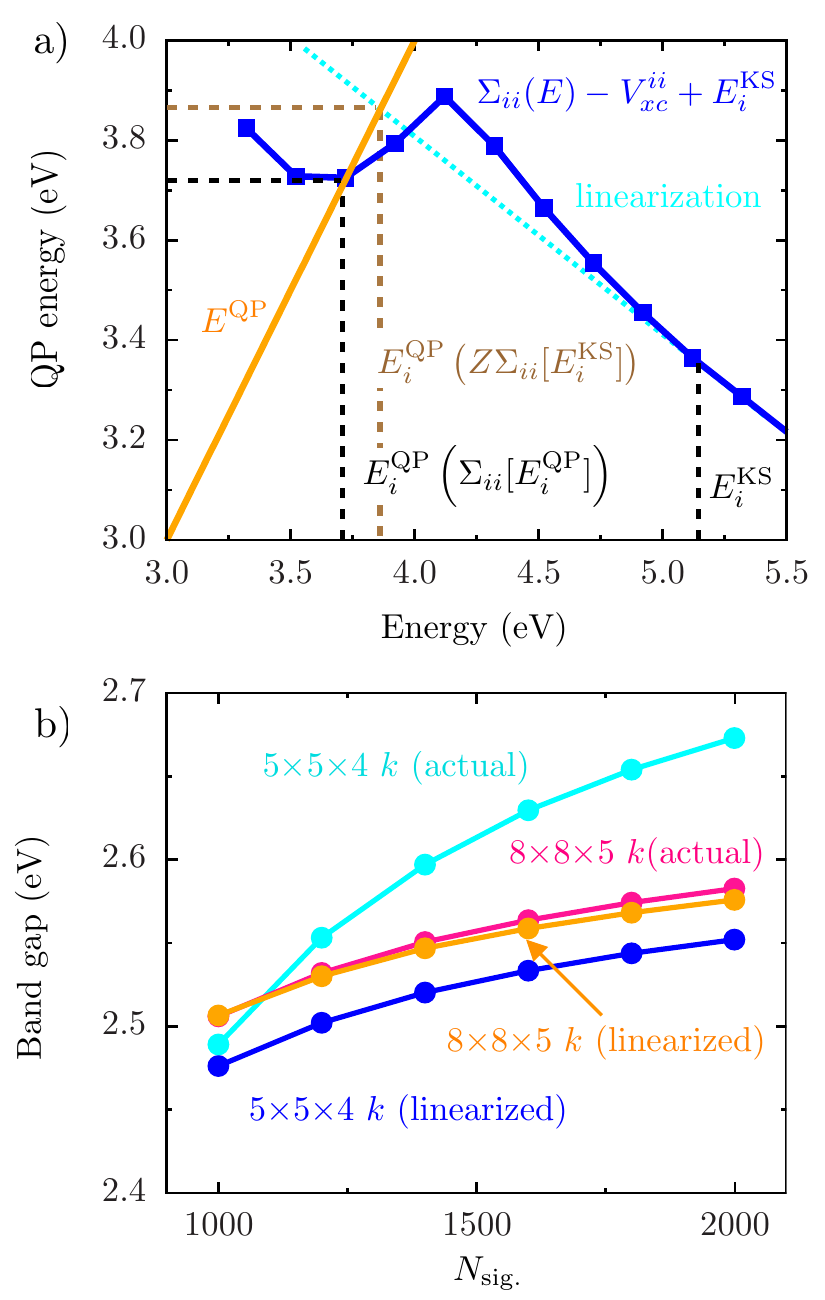}
\caption{Linearized vs. actual QP energies for ZnO.
a) QP energy for the VBM at the $\Gamma$ point. 
We show the actual self-energy, $E_i^{\qp}(\Sigma[E_i^{\qp}])$, and the linearized self-energy evaluated at the KS energy.
b) QP bandgap of ZnO. 
Two shifted k-point grids of $5\times5\times4$ and $8\times8\times5$ points are used.
The linearized and actual solutions disagree by more than $0.2$~eV for the coarser grid, and agree better than 50 meV for the finer grid of $8\times8\times5$ points. 
Here we use unconverged $GW$ parameters, as explained in the text.
}
\label{fig:linearized-ZnO}
\end{figure}
\begin{figure}[h]
\includegraphics{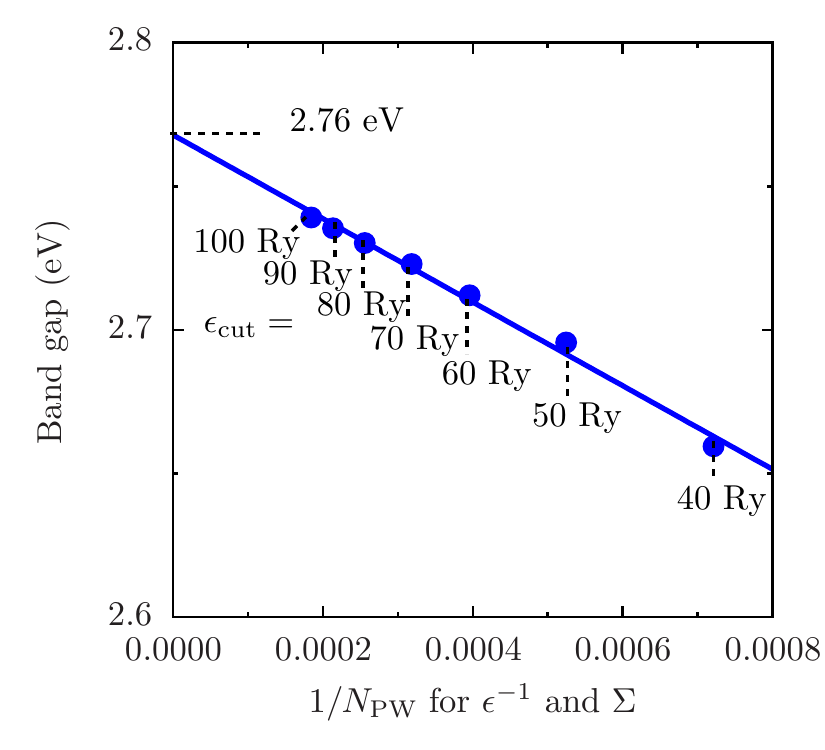}
\caption{Convergence of the bandgap of ZnO with respect to the plane-wave basis-set size.
The bandgap converges linearly with respect to $1/\Npw$. 
}
\label{fig:converged-ZnO}
\end{figure}
\begin{table}[ht]
\begin{ruledtabular}
\begin{tabular}{ccccc}
\multicolumn{5}{c}{ZnO QP bandgap (eV)}\\
Code & Potential	& Freq. & $E_g$	&	Ref. \\
\hline
        \bgw  	& NC-PP	& PPM-HL & 3.4	& \onlinecite{shih_quasiparticle_2010}\\
        \fullabinit 	& NC-PP	& PPM-HL & 3.6	& \onlinecite{stankovski_$g^0w^0$_2011}, \onlinecite{miglio_effects_2012}\\
        \textsc{Tombo}	& AE	& PPM-HL& 4.5	& \onlinecite{zhang_all-electron_2016}\\
    \fullabinit 	& NC-PP & PPM-HL	& 2.8	&
\onlinecite{larson_role_2013}\\ 
	\bgw 		& NC-PP	& PPM-HL$^*$	& 3.0	& \onlinecite{samsonidze_insights_2014}\\
	\fullabinit 	& NC-PP	& PPM-GN & 2.3	& \onlinecite{stankovski_$g^0w^0$_2011}, \onlinecite{miglio_effects_2012}\\
	\fullabinit 	& NC-PP	& PPM-GN & 2.6	& \onlinecite{berger_efficient_2012}\\
    &  AE	&	FF$^+$	& 2.4	& \onlinecite{usuda_all-electron_2002}\\
	\fullabinit 	& NC-PP	& FF-CD 	& 2.4	&  \onlinecite{stankovski_$g^0w^0$_2011}, \onlinecite{miglio_effects_2012}\\
    \textsc{Vasp} & PAW & FF-RA & 2.5 & \onlinecite{lim_angle-resolved_2012}\\
    & AE	& FF-CD	& 2.8	& \onlinecite{friedrich_band_2011,*friedrich_erratum:_2011}\\
            \textsc{Vasp}	&NC-PAW & FF-RA	&2.8 & \onlinecite{klimes_predictive_2014}\\
            \textsc{Tombo}	& AE	& FF$^\dagger$	& 2.8	& \onlinecite{zhang_all-electron_2016}\\
	\bgw    & NC-PP & FF-CD  & 2.8& This work\\
\end{tabular}
\end{ruledtabular}
\caption{Fundamental bandgap of \textrm{ZnO} within $G_0W_0$@LDA. 
The converged gap is extrapolated to the CBS, as detailed in the text.
The reported bandgaps using different codes and techniques are shown for comparison.
*~semicore electrons were excluded to calculate the ground-state density required to fit the PPM-HL parameters, see Ref.~\onlinecite{samsonidze_insights_2014}.
\label{table:zno}
$\dagger$ FF integration in the complex plane.
$+$ Frequency integration method based on the random-phase approximation~\cite{anisimov_strong_2000,usuda_all-electron_2002}.
}
\label{table:zno-gaps}
\end{table}

Historically, ZnO has been a challenging and controversial system for $GW$.
For ZnO, the $GW$ result is strongly affected by the slow convergence of the $\Sigma_c$ term~\cite{shih_quasiparticle_2010}.
Convergence issues are further aggravated when using PPMs~\cite{stankovski_$g^0w^0$_2011}, although these PPM-related issues may be partially remedied as illustrated in Ref.~\onlinecite{samsonidze_insights_2014}.
Here we only show results with FF methods and the PPM-GN (validated against FF references\cite{stankovski_$g^0w^0$_2011}). 
For more on the PPM approximation for ZnO, we refer the reader to Refs.~\onlinecite{stankovski_$g^0w^0$_2011,miglio_effects_2012,larson_role_2013,samsonidze_insights_2014}.
Other discrepancies in the $GW$ gap of ZnO arise from the use of incomplete basis-sets and different pseudopotentials, such as projector-augmented waves~\cite{klimes_predictive_2014}.
Due to these issues, the reported $G_0W_0$@LDA gaps with different approximations and codes range from 2.3 -- 4.5~eV (see Table~\ref{table:zno}).

We start by showing that the different codes used here agree on the gap of ZnO, for a given pseudopotential.
Again, although pseudopotential issues are not discussed here, we find that our results are insensitive (within 0.1~eV) to the choice of PPs tested in this work, as discussed in Appendix~\ref{appendix:pps}.
In Table~\ref{table:ZnO-spherical-cutoff}, we show underconverged QP energies for ZnO calculated using a spherical-cutoff scheme within $G_0W_0$@LDA. All ZnO results in Table V are computed at the same number of bands, dielectric matrix cutoffs, and k-point grid for the purposes of comparison. However, these parameters are underconverged.)
We use the GN method, FF-CD method with \fullabinit and \fullbgw, and the FF-RA method with \fullyambo.
We set $\ecuteps=30$~Ry, a Coulomb cutoff radius of 19.7177 Bohr, a
plasma frequency of	38.82 eV (for PPM-GN), a $\Gamma$-centered homogeneous grid of $5\times 5\times 4$ k-points and 34 bands, and show that the unconverged $GW$ gap of ZnO calculated with the different codes is consistent within 0.1~eV.

Linearizing the self-energy to the QP energy, especially when using coarse $k$-grids, can be inaccurate.
An illustration of the difference between the linearized and graphically-solved QP energies is given in Fig.~\ref{fig:linearized-ZnO}a. For the VBM, the linearized and graphical solutions can differ by $\sim 0.2$~eV; 
for an unconverged  set of $GW$ parameters ($5\times5\times4$~$k$ grid, $\ecuteps=40$~Ry and $\Neps=\Nsig=2000$), 
we find $E_i^{\qp}(\Sigma[E_i^{\qp}])=3.7$~eV and $E_i^{\qp}\left(Z\Sigma[E^{\ks}]\right)=3.9$~eV, where $E_i^{\ks}=5.3$~eV.
In Fig.~\ref{fig:linearized-ZnO}b we show the QP bandgap as a function of the number of bands used to evaluate $\Sigma$.
We use shifted grids of $5\times5\times4$ and $8\times8\times5$ k-point grids, $\Neps= 2000$ and $\ecuteps=40$~Ry.
Within the coarser grid the actual (blue dots) and linearized (cyan dots) solutions can disagree by more than $0.1$~eV due to features in $\Sigma(\omega)$, as shown in Fig.~\ref{fig:linearized-ZnO} (a). These features are smoothed out when using a finer grid, reducing the discrepancy associated with linearization.

Having demonstrated good agreement between different codes for ZnO QP energies, we then proceed to converge the gap of ZnO only with \fullbgw, excluding the other codes due to our limits on computational resources.
To accelerate the convergence with respect to k-points, we use a shifted grid, a common practice well-documented in the past\cite{sottile_parameter-free_2003}.
Using the finest grid of k-points (that is the $8\times8\times5$ grid), we proceed to converge the $\Neps$, $\Nsig$ and $\ecuteps$ by extrapolating to the CBS limit (see SI).
As shown in Fig.~\ref{fig:converged-ZnO}, the bandgap converges linearly with respect to $\Npw^{-1}$ and a relatively high $\ecuteps>80$~Ry is needed to assure convergence within 0.05~eV.
At convergence, we find the $G_0W_0$@LDA gap of ZnO is $2.8$~eV, in agreement with recent calculations, as shown in Table~\ref{table:zno-gaps}.

Finally, we compare our $G_0W_0$ bandgaps with the corresponding electronic gaps measured in photoemission experiments.
Here we use full-frequency $G_0W_0$ approaches (FF-CD or FF-RA).
Note that when comparing to experiment the lattice-renormalization effect should also be included~\cite{botti_strong_2013,antonius2014}, e.g., the measured/calculated zero-point renormalization~(ZPR) of silicon is 62--64~meV, 150 meV for TiO$_2$ and 156--164 meV for ZnO~\cite{cardona_isotope_2005,monserrat_correlation_2016}.
Our calculated indirect gap of 1.21--1.28 eV for silicon (without renormalization) is therefore in good agreement with the experimental gap of 1.17~eV~\cite{Kittel:ISSP}.
Our result is also in agreement with the seminal work of Ref.~\onlinecite{hybertsen_electron_1986}.
As mentioned above, since we neglect spin-orbit effects in this work, we do not compare the $GW$ bandstructure of gold to experiment.
Our calculated gap of 3.3~eV of rutile TiO$_2$ is also in good agreement with the experimental gap of $3.3\pm 0.5$~eV~\cite{tezuka_photoemission_1994,see_electronic_1994}.
On the other hand, our $GW$ gap of ZnO of 2.8~eV substantially underestimates the reported experimental gap of $\sim 3.6$~eV~\cite{tsoi_isotopic-mass_2006,alawadhi_effect_2007}.
This well-known shortcoming of standard $G_0W_0$ for ZnO is due to a deficient LDA starting point~\cite{lim_angle-resolved_2012}, and indicates the need for a more accurate starting point or self-consistent schemes.
This work reaches a consensus on the value of the $G_0W_0$ band-gaps of prototype systems, and hence facilitates future work studying beyond-standard $GW$ schemes to improve the accuracy of $GW$ when using a poor mean-field starting point.

\section{Conclusions}

In this work, we have revisited the $GW$ approximation for prototype systems with three representative plane-wave-based codes: \fullyambo, \fullabinit and \fullbgw.
Within certain choices of approximations and a given set pseudopotentials, the converged $GW$ QP energies calculated with the different codes agree within 0.1~eV, addressing long-standing controversies on the $GW$ results for difficult systems such as ZnO and rutile.

More specifically, we have studied the validity of approximations within one-shot $G_0W_0$ which can give rise to disagreement in $GW$ results between different codes, in particular the treatment of the Coulomb divergence, convergence, plasmon-pole model approximations, and scheme for capturing the full frequency dependence of $\Sigma$.
We have benchmarked different techniques to treat the Coulomb divergence, and identified several effective techniques, in particular an auxiliary-function method used in \fullabinit, the RIM in \fullyambo and the MC average in \fullbgw.
The latter was implemented in \fullabinit\ in this work.
We have provided new insights into the details of PPMs and their effect on $GW$ results, such as the treatment of unfulfilled PPM modes, which for some systems can lead to large deviations ($>0.5$~eV) from FF references.
We have shown that specific PPM techniques, when treated at the same level in the different codes, provide results in complete agreement independently of the code. Beyond the PPM approximation we have also shown that the FF-CD method implemented in \fullbgw\ provides results in agreement with FF  implementations in \fullabinit\ and \fullyambo. 
We highlight that QP energies predicted with the FF-CD method (in the complex plane) agree quantitatively with real-axis FF references, a numerical proof of the validity of the FF-CD. 

In summary, our work provides a framework for users and developers to validate and document the precision of new applications and methodological improvements relating to $GW$ codes. 

\section{Acknowledgments}

This work was supported by the Center for Computational Study of Excited State Phenomena in Energy Materials at the Lawrence Berkeley National Laboratory, which is funded by the U.S. Department of Energy, Office of Science, Basic Energy Sciences, Materials Sciences and Engineering Division under Contract No. DE-AC02-05CH11231, as part of the Computational Materials Sciences Program. 
This work is also supported by the Molecular Foundry through the U.S. Department of Energy, Office of Basic Energy Sciences under the same contract number.
We acknowledge the use of computational resources at the National Energy Research Scientific Computing Center (NERSC).
We also acknowledge the use of HPC resources from GENCI-CCRT-TGCC (Grants No. 2014-096018).
F.B. acknowledges the Enhanced Eurotalent program and the France Berkeley Fund for supporting his sabbatical leave in UC Berkeley. D.V., A.F. and A.M.  acknowledge support from European Union H2020-INFRAEDI-2018-1 programme under grant agreement No. 824143 project ``MaX - materials at the exascale''. 
AM acknowledges support from European Union H2020-INFRAIA-2015-1 programme under grant agreement No. 676598 project ``Nanoscience Foundries and Fine Analysis - Europe''.
O.K.O. and D.O'.R. acknowledge the support of Trinity College Dublin's Studentship Award
and School of Physics. Their work was supported by TCHPC (Research IT, Trinity College Dublin), where calculations were performed on the Lonsdale cluster funded through grants from Science Foundation Ireland, and on the Kelvin cluster funded through grants from the Irish Higher Education Authority through its PRTLI program.
D.V. and A.F. acknowledge PRACE for awarding access to resource Marconi based in Italy at CINECA.
M.J.v.S. and G.-M.R. are grateful to F.R.S.-FNRS for financial support through the PDR Grants T.1031.14 (HiT4FiT). They also thank the C{\'E}CI facilities funded by F.R.S.-FNRS (Grant No. 2.5020.1) and Tier-1 supercomputer of the F{\'e}d{\'e}ration Wallonie-Bruxelles funded by the Walloon Region (Grant No. 1117545). 

\appendix
\section{The choice of pseudopotential for $GW$}
\label{appendix:pps}

\begin{table}[b!]
\begin{ruledtabular}
\begin{tabular}{crlccc}
\multicolumn{6}{c}{Pseudopotentials for TiO$_2$}\\
\hline
PP   & \multicolumn{2}{c}{PP radii} & $E_\textrm{\scriptsize cut.}$ & DFT gap & $GW$ gap\\
type & \multicolumn{2}{c}{(Bohr)}   & (Ry)                          & (eV)    & (eV)\\
\hline 
\\
FHI\footnote{
FHI PPs: Ti is defined in the OPIUM-v3.8 user guide\cite{opium-site} and O is from the FHI98 library~\cite{fuchs_ab_1999,abinit-site}.
}
& Ti:& s 1.48, p 1.62, d 1.70 & \multirow{1}{*}{60}  & \multirow{1}{*}{1.78} & \multirow{1}{*}{3.12}\\
\\
\multirow{2}{*}{
HGH\footnote{HGH PPs from Refs.~\onlinecite{hartwigsen_relativistic_1998,abinit-site}.}}
& Ti:& s 0.34, p 0.24, d 0.24 & \multirow{2}{*}{280} & \multirow{2}{*}{1.88} & \multirow{2}{*}{3.23}\\
     &  O:& s 0.22, p 0.21 & & &\\
\\
\multirow{2}{*}{
PD\footnote{PD PPs from pseudo-dojo-v.2~\cite{pseudo-dojo, VANSETTEN201839}.}}
& Ti:& s 1.35, p 1.30, d 1.65 & \multirow{2}{*}{60}  & \multirow{2}{*}{1.88} & \multirow{2}{*}{3.23} \\
     &  O:&  s 1.25, p 1.35& & &\\
\\
\end{tabular}
\end{ruledtabular}
\caption{Testing norm-conserving pseudopotentials for rutile.
For each PP type we show the radii per angular momentum (s, p or d), the plane-wave energy-cutoff ($E_\textrm{\scriptsize cut.}$) (see text), and the corresponding DFT and  $GW$ gap of rutile.
We use $G_0W_0$ PPM-GN with a DFT-PBE starting point, at unconverged $GW$ parameters (see text).
}
\label{table:rutile-pseudos}
\end{table}

\begin{table}[thb]
\vspace{0.1in}
\begin{ruledtabular}
\begin{tabular}{crlccc}
\multicolumn{6}{c}{Pseudopotentials for ZnO}\\
\hline
PP   & \multicolumn{2}{c}{PP radii} & $E_\textrm{\scriptsize cut.}$ & DFT gap & $GW$ gap\\
type & \multicolumn{2}{c}{(Bohr)}   & (Ry)                          & (eV)    & (eV)\\
\hline
\\
\multirow{2}{*}{FHI\footnote{FHI PPs from Ref.~\onlinecite{stankovski_$g^0w^0$_2011}.}
}  & Zn:& s 0.80, p 0.80, d 0.80 & \multirow{2}{*}{300}  & \multirow{2}{*}{0.67} & \multirow{2}{*}{2.76}\\
     & O: & s 1.20, p 1.20 & & & \\
\\
\multirow{2}{*}{RRKJ\footnote{RRKJ\cite{rappe_optimized_1990} PPs from Ref.~\onlinecite{samsonidze_insights_2014}.}
}  & Zn:& s 1.00, p 1.00, d 0.85 & \multirow{2}{*}{300}  & \multirow{2}{*}{0.73} & \multirow{2}{*}{2.87}\\
     & O: & s 1.10, p 1.10 & & & \\
\\
\multirow{2}{*}{HGH\footnote{HGH PPs from Refs.~\onlinecite{hartwigsen_relativistic_1998,abinit-site}.}
}  & Zn:& s 0.40, p 0.53, d 0.25 & \multirow{2}{*}{300} & \multirow{2}{*}{0.73} & \multirow{2}{*}{2.90}\\
     &  O:& s 0.22, p 0.21 & & &\\
\\
\multirow{2}{*}{PD\footnote{PD PPs from pseudo-dojo-v.2~\cite{pseudo-dojo, VANSETTEN201839}.}
} & Zn:& s 1.35, p 1.65, d 1.85 & \multirow{2}{*}{60}  & \multirow{2}{*}{0.78} & \multirow{2}{*}{2.82} \\
     &  O:&  s 1.25, p 1.35& & &\\
\\
\multirow{2}{*}{PD\footnote{PD PPs generated with the ONCVP code~\cite{hamann_optimized_2013}.}
} & Zn:& s 0.80, p 0.80, d 0.60 & \multirow{2}{*}{500}  & \multirow{2}{*}{0.74} & \multirow{2}{*}{2.84} \\
     &  O:&  s 0.80, p 0.80& & &\\
\\
\end{tabular}
\end{ruledtabular}
\caption{Sensitiveness of the $GW$ gap of ZnO with respect to the choice of PPs.
Same as TiO$_2$ in Table~\ref{table:rutile-pseudos}.
We use $G_0W_0$ FF-CD with a DFT-LDA starting point at unconverged $GW$ parameters (see text).
Note that the $GW$ gaps of ZnO shown in this table agree with the converged gap ($=2.8$~eV) due to spurious cancellation of errors. 
}
\label{table:zno-pseudos}
\end{table}

In this appendix, we study the variation of the bandgap with respect to the {\it choice of pseudopotential} for TiO$_2$ and ZnO.
We emphasize that the validation of pseudopotentials for $GW$ requires all-electron references and is beyond the scope of the present manuscript.
In Table~\ref{table:rutile-pseudos} we show the $G_0W_0$ direct gap of rutile calculated with different choices of pseudopotentials.
We use a DFT-PBE starting point from \fullabinit\ and consider norm-conserving PPs of the Fritz Haber Institute~(FHI)\cite{fuchs_ab_1999}, Optimized Norm-Conserving Vanderbilt~(ONCV)\cite{hamann_optimized_2013} and Hartwigsen-Goedecker-Hutter~(HGH)\cite{hartwigsen_relativistic_1998} kinds.
The configuration of choice for Ti is [Ne]3s$^2$3p$^6$3d$^2$4s$^2$ (including semicore states), and [He]2s$^2$2p$^6$ for O.
We only use PPs available in the literature (see Table~\ref{table:rutile-pseudos}).
Note that the HGH and PD PPs contain non-local core corrections~(NLCC), which are subtracted from $\Sigma$ when calculating the QP energies.
In the table, we show the energy cutoff required to converge the DFT total energy per atom to 0.01~eV and the PP radii, which can be taken as a measure of the PP ``hardness''.
Here we use \fullbgw\ to compute the $G_0W_0$ direct gap of rutile using a set of under-converge parameters for $GW$: $\Neps=\Nsig=2000$, $\ecuteps=20$~Ry, the MC avg. technique and a $\Gamma$-centered homogeneous grid of $6\times6\times10$~k-points.
Importantly, the $GW$ gaps correponding to different PP types agree within 0.1~eV, indicating a small dependence of the gap of rutile with the choice of PPs used here.

We now study the sensitiveness of the $GW$ results with respect to the choice of pseudopotential for ZnO.
In Table~\ref{table:zno-pseudos} we show the QP gap of ZnO calculated with $G_0W_0$@LDA using different PPs.
The configuration of choice for Zn is [Ne]3s$^2$3p$^6$3d$^{10}$4s$^2$ (including semicore states), and [He]2s$^2$2p$^6$ for O.
As in the TiO$_2$ case, some of the HGH and PSP8 PPs considered here contain NLCCs.
We also show the minimum kinetic energy cutoff for the plane-wave expansion to converge the DFT gap within $0.05$~eV, and the corresponding DFT-LDA and $GW$ gaps.
Here we use \fullbgw, the FF-CD method with 20 imaginary frequencies, an uniform sampling of real frequencies spaced by 0.25~eV from 0 to 6~eV, the modified static-reminder method of Ref.~\onlinecite{deslippe_coulomb-hole_2013} and unconverged $GW$ parameters: $\ecuteps=30$~Ry, $\Nsig=\Neps=500$ .
For ZnO the $GW$ and DFT gaps change little, by up to 0.14 and 0.1~eV respectively, with the different choices of PPs.
Therefore, the results for ZnO and TiO$_2$  presented in this manuscript are negligibly affected by the choice of PPs.

\bibliography{main}

\end{document}


\newcommand{\ecuteps}{\epsilon_{\scriptsize\textrm{cut}}}
\newcommand{\bgw}{\textsc{BGW}\xspace}
\newcommand{\abinit}{\textsc{ABI}\xspace}
\newcommand{\yambo}{\textsc{YMB}\xspace}
\newcommand{\fullbgw}{\textsc{BerkeleyGW}\xspace}
\newcommand{\fullabinit}{\textsc{Abinit}\xspace}
\newcommand{\fullyambo}{\textsc{Yambo}\xspace}
\newcommand{\exx}{E_{\scriptsize\textsc{EXX}}}
\newcommand{\qp}{\textrm{\scriptsize{QP}}}
\newcommand{\ks}{\textrm{\scriptsize{KS}}}
\newcommand{\unfmode}{\omega^{\scriptsize\textrm{unf.}}}
\newcommand{\Neps}{N_{\scriptsize\textrm{eps.}}}
\newcommand{\Nsig}{N_{\scriptsize\textrm{sig.}}}
\newcommand{\Npw}{N_{\scriptsize\textrm{PW}}}

\title{Supplemental Information for ``Reproducibility in the $G_0W_0$ Calculations for Solids''}

\author{Tonatiuh Rangel}
\email{trangel@lbl.gov}
\affiliation{Molecular Foundry, Lawrence Berkeley National Laboratory, Berkeley, California 94720, United States}
\affiliation{Department of Physics, University of California at Berkeley, California 94720, United States}
\author{Mauro Del Ben}
\affiliation{Computational Research Division, Lawrence Berkeley National Laboratory, Berkeley, California, 94720, United States}
\author{Daniele Varsano}
\affiliation{Centro S3, CNR-Istituto Nanoscienze, I-41125 Modena, Italy}
\affiliation{European Theoretical Spectroscopy Facility (ETSF)}
\author{Gabriel Antonius}
\affiliation{Department of Physics, University of California at Berkeley, California 94720, United States}
\affiliation{Materials Sciences Division, Lawrence Berkeley National Laboratory, Berkeley, California 94720, United States}
\affiliation{Département de Chimie, Biochimie et Physique, Institut de recherche sur l’hydrogène, Université du Québec à Trois-Rivières, Qc, Canada}
\author{Fabien Bruneval}
\affiliation{DEN, Service de Recherches de Métallurgie Physique, Université Paris-Saclay, CEA, F-91191 Gif-sur-Yvette, France}
\affiliation{Molecular Foundry, Lawrence Berkeley National Laboratory, Berkeley, California 94720, United States}
\affiliation{Materials Sciences Division, Lawrence Berkeley National Laboratory, Berkeley, California 94720, United States}
\author{ Felipe H. {da Jornada}}
\affiliation{Department of Physics, University of California at Berkeley, California 94720, United States}
\affiliation{Materials Sciences Division, Lawrence Berkeley National Laboratory, Berkeley, California 94720, United States}
\author{ Michiel J. {van Setten}}
\affiliation{Institute of Condensed Matter and Nanoscience (IMCN), Universit\'e catholique de Louvain, 1348 Louvain-la-Neuve, Belgium}
\affiliation{European Theoretical Spectroscopy Facility (ETSF)}
\affiliation{IMEC, Kapeldreef 75, 3001 Leuven, Belgium}
\author{Okan K. Orhan}
\affiliation{School of Physics, Trinity College Dublin, The University of Dublin, Dublin 2, Ireland}
\author{ David D. {O'Regan}}
\affiliation{School of Physics, Trinity College Dublin, The University of Dublin, Dublin 2, Ireland}
\author{Andrew Canning}
\affiliation{Computational Research Division, Lawrence Berkeley National Laboratory, Berkeley, California, 94720, United States}
\author{Andrea Ferretti}
\affiliation{Centro S3, CNR-Istituto Nanoscienze, I-41125 Modena, Italy}
\affiliation{European Theoretical Spectroscopy Facility (ETSF)}
\author{Andrea Marini}
\affiliation{Istituto di Struttura della Materia of the National Research Council, Via Salaria Km 29.3, I-00016 Montelibretti, Italy}
\affiliation{European Theoretical Spectroscopy Facility (ETSF)}
\author{Gian-Marco Rignanese}
\affiliation{Institute of Condensed Matter and Nanoscience (IMCN), Universit\'e catholique de Louvain, 1348 Louvain-la-Neuve, Belgium}
\affiliation{European Theoretical Spectroscopy Facility (ETSF)}
\author{Jack Deslippe}
\affiliation{NERSC, Lawrence Berkeley National Laboratory, Berkeley, California 94720, United States}
\author{Steven G. Louie}
\affiliation{Department of Physics, University of California at Berkeley, California 94720, United States}
\affiliation{Materials Sciences Division, Lawrence Berkeley National Laboratory, Berkeley, California 94720, United States}
\author{Jeffrey B. Neaton}
\affiliation{Molecular Foundry, Lawrence Berkeley National Laboratory, Berkeley, California 94720, United States}
\affiliation{Department of Physics, University of California at Berkeley, California 94720, United States}
\affiliation{Kavli Energy Nanosciences Institute at Berkeley, Berkeley, California 94720, United States}


\begin{abstract}
In this Supplemental Information we show convergence studies on the QP energies of silicon, gold, TiO$_2$ and zinc oxide.
We show the technical parameters used in our $GW$ calculations.
We describe a simple strategy to compare the results from different $GW$ codes.
Finally, we study the sensitiveness of the $GW$ gaps to the choice of pseudopotential. 
\end{abstract}

\maketitle

\section{Technical details}
\label{sect:technical-details}

In Table~\ref{Table:tech-details} we summarize all parameters used in our $GW$ calculations for Tables and Figures in the main manuscript.  
\begin{table*}[ht]
\begin{ruledtabular}
\begin{tabular}{lccccccc}
$k$-point grid & $\ecuteps$ & $\Neps$ & $\Nsig$ & PP & St. point & Sph. cut. & $\Omega_p$ \\ 
\hline
\multicolumn{6}{l}{\it Silicon}\\
\multicolumn{6}{l}{Table I of main manuscript}\\
$12\times12\times12$, unshifted & 40 Ry & 300 & 300 & FHI$^a$ & PBE & No & 16.60 eV\\
\multicolumn{6}{l}{Fig. 1 of main manuscript}\\
$4\times4\times4$, unshifted & 20 Ry &  & 4 & FHI$^a$ & PBE & Yes & 16.60 eV\\
\hline
\multicolumn{6}{l}{\it Gold}\\
\multicolumn{6}{l}{Table II of main manuscript}\\
$16\times16\times16$, unshifted & 32 Ry & 400 & 400 & PD$^b$ & PBE & No & 38.88 eV\\
\hline
\multicolumn{6}{l}{\it Rutile}\\
\multicolumn{6}{l}{Table III of main manuscript}\\
$6\times6\times10$, unshifted & 20 Ry & 100 & 100 & FHI$^c$ & PBE & Yes & 32.56 eV\\
\multicolumn{6}{l}{Table IV of main manuscript}\\
$6\times6\times10$, unshifted & \multicolumn{3}{c}{extrapolated to CBS limit$^d$}  & FHI$^c$ & PBE & No & 32.56 eV\\
\hline
\multicolumn{6}{l}{\it Zinc Oxide}\\
\multicolumn{6}{l}{Table V of main manuscript}\\
$5\times5\times4$, unshifted & 30 Ry & 34 & 34 & FHI$^e$ & LDA & Yes & 38.82 eV\\
\multicolumn{6}{l}{Fig. 3a of main manuscript}\\
$5\times5\times4$ & 40 Ry & 2000 & 2000 & FHI$^e$ & LDA & No & \\
\multicolumn{6}{l}{Fig. 3b of main manuscript}\\
Several grids$^{f,g}$ & 40 Ry & 2000 & 2000 & FHI$^e$ & LDA & No & \\
\multicolumn{6}{l}{Fig. 4 of main manuscript}\\
$8\times8\times5$, shifted$^g$ & \multicolumn{3}{c}{extrapolated to CBS limit$^h$} & FHI$^e$ & LDA & No & \\
\multicolumn{6}{l}{Table VI of main manuscript}\\
$8\times8\times5$, shifted$^g$ & \multicolumn{3}{c}{extrapolated to CBS limit$^h$} & FHI$^e$ & LDA & No & \\

\end{tabular}
\flushleft
$^a$ FHI98 PP~\cite{fuchs_ab_1999} from the ABINIT web site.\\
$^b$ PD PP from pseudo-dojo-v.2~\cite{pseudo-dojo, VANSETTEN201839}.\\
$^c$ FHI PPs: Ti is defined in the OPIUM-v3.8 user guide\cite{opium-site} and O is from the FHI98 library~\cite{fuchs_ab_1999,abinit-site}.\\
$^d$ Number of bands and $\ecuteps$ extrapolated to the complete basis set (CBS) limit, see Section~\ref{rutile}.\\ 
$^e$ FHI PPs from Ref.~\onlinecite{stankovski_$g^0w^0$_2011}.\\
$^f$ Shifted and unshifted grids of $5\times5\times4$ and $8\times8\times5$ $k$-points.\\ 
$^g$ {\it Shifted grids} are only used calculate the dielectric constant.\\
$^h$ Extrapolation to the CBS limit, read Section D of main manuscript.\\
\end{ruledtabular}
\caption{
Parameters used in our $GW$ calculations.
We report our choice of $k$-point grid, energy cutoff for the screening ($\ecuteps$), number of bands used to calculate the dielectric function ($\Neps$) and the $GW$ self-energy ($\Nsig$), pseudopotential (PP), DFT starting point (St. point), whether a spherical cutoff was used (Sph. cut.).
Additionally, a plasma frequency ($\Omega_p$) is shown when referring to PPM-GN calculations. 
}
\label{Table:tech-details}
\end{table*}

\section{Strategy to validate codes}
\label{sect:strategy}

Comparing the numerical accuracy of different $GW$ codes is a complex task due to the various approximations used in the practice, such as the degree of convergence, the treatment of the Coulomb singularity and frequency dependence of $\Sigma(\omega)$ and the choice of pseudopotantials. 
Nevertheless, we can isolate the different sources of discrepancy following a {\it simple strategy}: 
\begin{enumerate}
\item We start with exactly the same DFT input (wavefunctions and eigenvalues).
\item We avoid the Coulomb singularity by using the spherical-cutoff scheme; and we use the same frequency integration scheme, number of bands, and $k$-points in BZ integrations.
\item We explore $k$-point convergence, which is sensitive to the method used to treat the Coulomb singularity (see above).
\item We converge $\ecuteps$ simultaneously with the number of bands used to evaluate $\epsilon$ and $\Sigma$ ($\Neps$ and $\Nsig$ respectively).
\end{enumerate}

Below, we describe in more detail our application of this strategy to validate $GW$ codes for bulk crystalline Si, Au, TiO$_2$, and ZnO.

In TABLE~\ref{table:spherical-truncation}, we show the $GW$ corrections to the valence band maximum~(VBM) and conduction band minimum~(CBM) at the $\Gamma$ point for bulk silicon.
We show the corrections calculated with/without the spherical-cutoff technique and using the three different codes used here \fullabinit, \fullbgw and \fullyambo.
We set $\ecuteps=20$~Ry, a grid of $4\times4\times4$~k-points and 10 empty states; with these settings, the $GW$ corrections are underconverged but the level of accuracy is the same for the different codes, facilitating quantitative comparison. When using the spherical-cutoff technique (see last two columns in TABLE~\ref{table:spherical-truncation}), we obtain the same QP energies with different codes.
This result shows that by disentangling the sources of discrepancy with our simple strategy, we can effectively compare output of different $GW$ codes. 

\begin{table}[t]
\begin{ruledtabular}
\begin{tabular}{l@{\hskip 0.1in}cc@{\hskip 
0.2in}cc}
\multicolumn{5}{c}{Unconverged $GW$ corrections for silicon at $\Gamma$ in eV}\\
\hline
\multicolumn{5}{c}{Valence band maximum}\\
& \multicolumn{2}{c}{No cutoff} 
& \multicolumn{2}{c}{Spherical cutoff} 
\\
& $\Sigma_x$ & $\Sigma_c$
& $\Sigma_x$ & $\Sigma_c$\\
\fullbgw    & -12.746 & 2.030 & -13.049 & 0.969 \\
\fullabinit & -13.057 & 2.178 & -13.049 & 0.969 \\
\fullyambo & -12.630 & 1.975 & -13.050 & 0.969\\
\hline
\multicolumn{5}{c}{Conduction band minimum}\\
& \multicolumn{2}{c}{No cutoff} 
& \multicolumn{2}{c}{Spherical cutoff} 
\\
& $\Sigma_x$ & $\Sigma_c$
& $\Sigma_x$ & $\Sigma_c$\\
\fullbgw    & -5.640 & -2.946 & -5.709 & -1.984\\
\fullabinit & -5.640 & -3.094 & -5.709 & -1.984 \\
\fullyambo  & -5.640 & -2.890 & -5.709& -1.984\\
\end{tabular}
\end{ruledtabular}
\caption{$GW$ corrections to the VBM/CBM of silicon with low convergence parameters (read text).
The unconverged $GW$ corrections are identical for different codes when using the spherical-cutoff technique (last two columns), illustrating how to cheaply evaluate the numerical accuracy of $GW$ codes.
}
\label{table:spherical-truncation}
\end{table}

\section{Gold}
\begin{figure}
\includegraphics{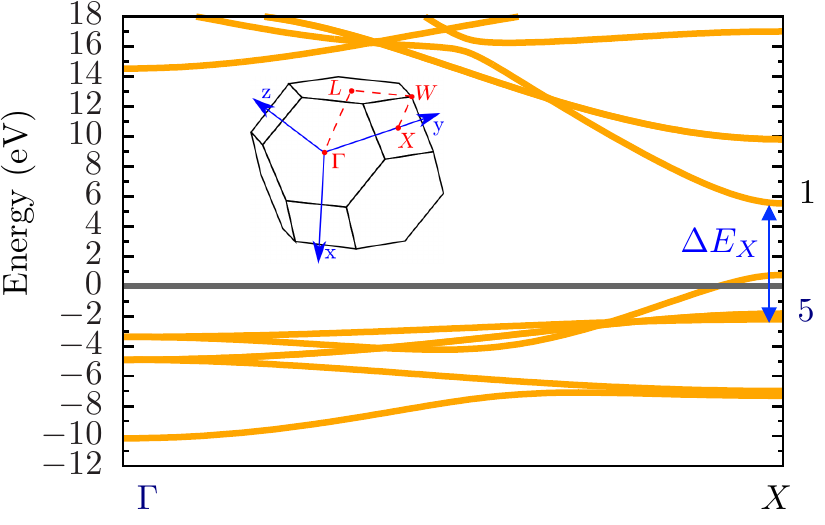}
\caption{Scalar-relativistic bandstructre of gold calculated with DFT-PBE. 
The Brillouin zone is shown as an inset.}
\label{fig:gold-bs}
\end{figure}
\begin{figure}
\includegraphics{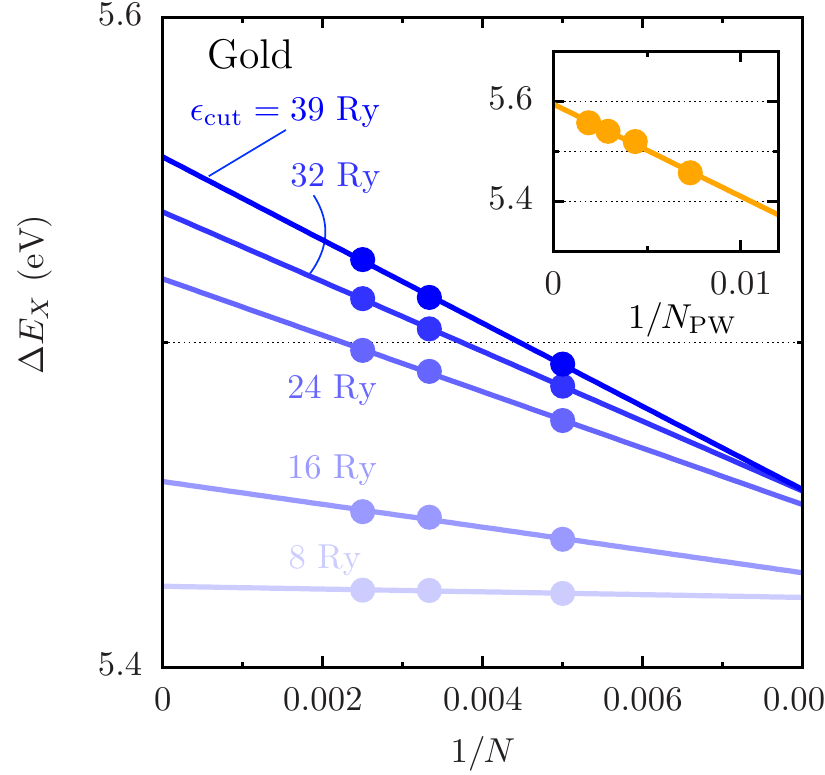}
\caption{Energy bandgap ${\Delta}E_X$, defined in FIG.~\ref{fig:gold-bs}, of gold.
The bandgap for different values of $\ecuteps$ and bands $N$ (for both $\Sigma_c$ and $\epsilon$) is linearly extrapolated to $N\rightarrow \infty$ (blue lines).
Inset: the resulting gaps are then extrapolated to the CBS limit ($\Npw \rightarrow \infty$).
}
\label{fig:gold-conv}
\end{figure}
\begin{figure}
\includegraphics[scale=0.95]{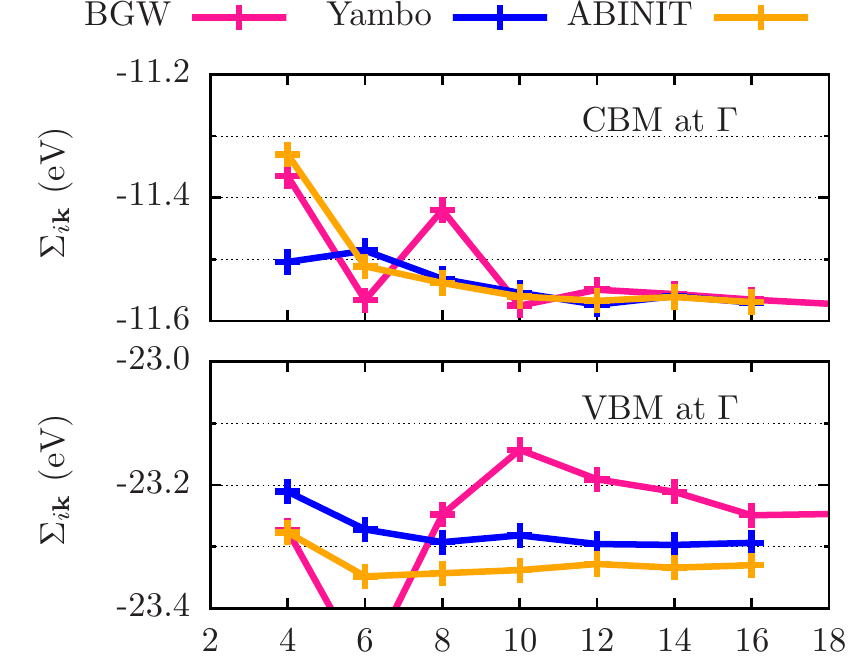}
\caption{Convergence of the $GW$ self-energy of gold.
We show $\Sigma_{i\mathbf{k}}$ matrix elements for $\mathbf{k}=\Gamma$ and $i=$VBM/CBM.
We consider uniform $k$-point grids of $N_k\times N_k \times N_k$ $k$ points.
The codes used here implement particular sets of approximations to treat metals (read text).
}
\label{fig:gold-k-points}
\end{figure}

Bulk gold is a more complex case for $GW$ due to approximations to the metal dielectric and the partial band occupations at the Fermi energy.
As mentioned in the manuscript, we study the $GW$ corrections to the scalar relativistic bandstructure of gold, shown in FIG.~\ref{fig:gold-bs}.
Notice that this should not be compared to experiment, as the strong spin-orbit effects are neglected.

We first converge the number of bands and planewaves used for the dielectric function and $\Sigma$, using a fixed $k$-point sampling of $8\times8\times8$ $k$-points, the PPM-GN, and the \fullabinit\ code.
The linear extrapolation to an infinite number of bands and planewaves is done in a two-step procedure.
First, for a fixed $\ecuteps$, we extrapolate the number of bands $N$ for $\Sigma$ and $\epsilon$ to $N \rightarrow \infty$ (see FIG.~\ref{fig:gold-conv}).
Second, the resulting gaps are linearly extrapolated to the complete basis set limit~(CBS) (see the inset). 
The choice of linear extrapolation is motivated by recent work\cite{klimes_predictive_2014}, and here the fit is satisfactory with a standard deviation of $<0.005$~eV.
We find an extrapolated gap at $X$ ($X_5 \rightarrow X_1$) of $5.59$~eV.
Setting this gap as the reference, we find that the set of parameters $\ecuteps=32$~Ry and $N=400$ are sufficient to converge the gap within $<0.1$~eV.


\section{T\MakeLowercase{i}O$_2$}
\label{rutile}

%
\begin{figure}[h]
\includegraphics[scale=0.95]{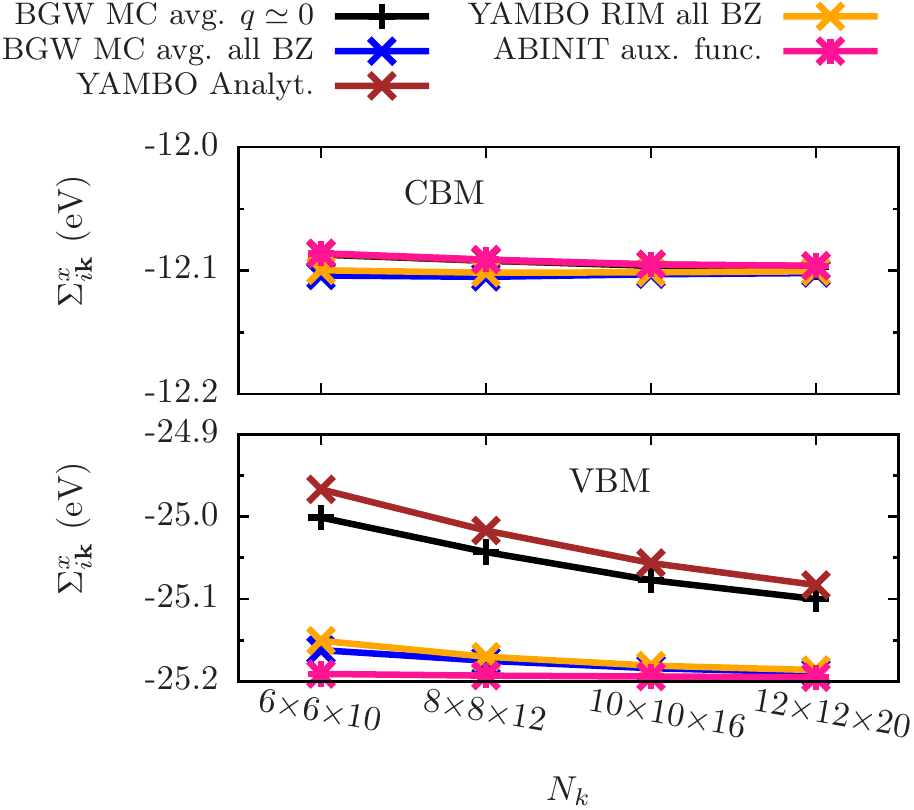}
\caption{Convergence of the $\Sigma^x_{i\mathbf{k}}$ matrix elements with respect to the $k$-point mesh for rutile.
Shown are the matrix elements for the $\Gamma$ point and the VBM/CBM.
Several codes and methods are considered (read text).
}
\label{fig:tio2-sigx}
\end{figure}
\begin{figure}[h]
\includegraphics{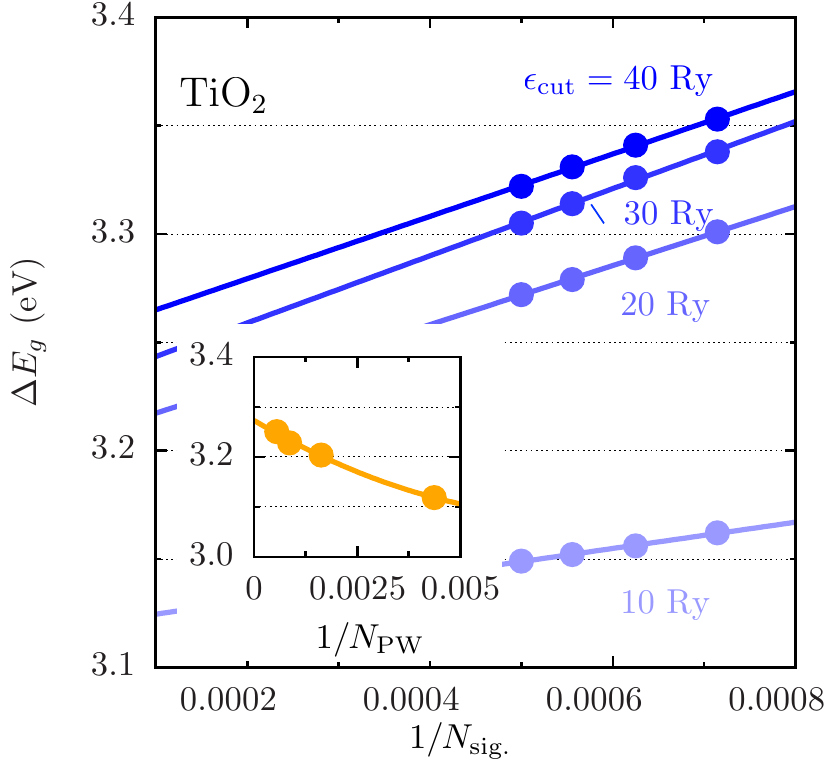}
\caption{Fundamental bandgap ${\Delta}E_g$ of rutile.  
The bandgap for different values of $\ecuteps$ and $\Nsig$ is extrapolated to the limit of an infinite number of bands (blue lines).
Here we fix $\Neps=2000$.
Inset: the resulting gaps are then extrapolated to the limit of an infinite number of plane-waves $N_\textrm{\scriptsize PW}$ in the basis set.
}
\label{fig:tio2-conv}
\end{figure}

For rutile Ti$O_2$, we first study the convergence of $\Sigma$ with respect to the number of bands and cutoff used for the $GW$ sums. 
In FIG.~\ref{fig:tio2-sigx}, we show $\Sigma^x_{i\mathbf{k}}$ matrix elements as a function of the $k$-point mesh.
Consistent with the silicon case (see main manuscript), some methods, such as the \fullyambo\ analytic method~(brown lines) and the \fullbgw\ MC avg. on $q\simeq0$~(black lines) can exhibit a slow convergence rate.
Importantly, with a mesh of $6\times6\times10$ $k$-points or denser, when using the RIM~(orange lines) or MC average~(blue lines) over the entire BZ, or the auxiliary function of Carrier~(pink lines), all codes converge $\Sigma^{x}_{i\mathbf{k}}$ for the VBM/CBM within $<0.1$~eV.
Moreover, in TABLE III and IV of the main manuscript, we show that the $GW$ codes used in this work agree within 0.1~eV in the predicted gap of rutile.

We now converge the $GW$ parameters with \fullbgw. 
In FIG.~\ref{fig:tio2-conv}, we show the gap of rutile for given values of $\ecuteps$ and $\Nsig$, fixing $\Neps$ to $2000$ (blue dots).
For each value of $\ecuteps$, we linearly extrapolate the gaps with $1/\Nsig$ to $\Nsig \rightarrow \infty$ (blue lines). 
The resulting gaps are then used to further extrapolate the gap to the CBS limit with $1/\Npw$;
a linear extrapolation results in a gap of $3.26 \pm 50$~meV,
and the accuracy is increased with a quadratic extrapolation resulting in $\Delta E_g= 3.27 \pm 10$~meV (see the inset). 
With this procedure, or alternatively by extrapolating simultaneously the parameters ($\ecuteps$, $\Neps$, and $\Nsig$) via Eqn.~7 of the main text, 
\begin{align}
f(\Neps, & \Npw, \Nsig) = \nonumber\\
 & \left(\frac{a_1}{\Neps} + b_1  \right)
   \left( \frac{a_2 }{\Npw}  + b_2 \right)
   \left(\frac{a_3  }{\Nsig} + b_3 \right),
\label{eq:extrapolation}
\end{align}
we obtain a converged $GW$ gap of rutile of $3.3$~eV.

\section{Z\MakeLowercase{n}O}
\label{sec:ZnO}

We show that the $GW$ codes considered here can predict the same QP energies for ZnO within a given choice of PPs and frequency-integration method.
In TABLE V of the main manuscript we show the VBM, CBM and fundamental gap of ZnO calculated with different codes and using a spherical-cutoff technique.
As our aim is to compare output of different codes, we reduce computational cost and use under-converged $GW$ parameters: $\ecuteps = 80$~Ry, $5\times5\times4$~$k$-points and 34 bands to evaluate $\Sigma$. 
As expected, we find a negligible deviation for the PPM (GN) with respect to FF-integration, in agreement with Refs.~\onlinecite{stankovski_$g^0w^0$_2011,miglio_effects_2012,larson_role_2013}. 
Remarkably, the QP energies obtained with different codes agree within $0.03$~eV.

Next, we converge the QP gap of ZnO with \fullbgw.
Here we use an LDA starting point, a shifted grid of $8\times8\times5$~$k$-points for $\epsilon$ and $\Sigma$, and FF-CD with the same frequency grid as above. 
We then extrapolate the fundamental gap of ZnO, similar to previous cases.
First, we calculate the gap for fixed values of $\ecuteps$ and extrapolating to $\Nsig \rightarrow \infty$ with $1/\Nsig$ (see FIG.~3 of main text);
here we set $\Neps=2000$.
Second, the gap is extrapolated to an infinit basis-set size with $1/\Npw$ (see FIG.~4 of main text).
The converged gap of ZnO with this procedure is 2.76~eV.
Alternatively, we can extrapolate the $GW$ parameters simultaneously via Eqn.~\ref{eq:extrapolation},
where for ZnO we find $a_1=-19.1213$, $a_2=-58.4158$, $a_3=-96.9096$, $b_1=1.3356$, $b_2=1.4735$ and $b_3=1.4127$ (in eV).
With this alternative procedure the extrapolated bandgap is $b_1b_2b_3=2.78$~eV, coinciding with the previously extrapolated gap (with a deviation of only 0.02~eV). 

\bibliography{SI}